\documentclass[11pt]{article} 
\usepackage{url}
\usepackage{latexsym}
\usepackage{multirow}
\usepackage{graphicx}
\graphicspath{}
\usepackage{booktabs}
\usepackage{float}
\usepackage{listings}
\usepackage[margin=1in]{geometry}
\usepackage{amsmath}
\usepackage{datetime}
\usepackage[toc,page]{appendix}

\usepackage{sidecap}
\sidecaptionvpos{figure}{c}

\newcommand{\bigCI}{\mathrel{\text{\scalebox{1.07}{$\perp\mkern-10mu\perp$}}}}

\usepackage{fancyhdr}
\newcommand\shorttitle{Predicting Preventable Hospital Readmissions with Causal Machine Learning}
\newcommand\authors{Marafino, Schuler, Liu, Escobar, and Baiocchi}

\usepackage[utf8]{inputenc}

\usepackage{titlesec}
\titleformat*{\section}{\large \centering \bfseries}
\titleformat*{\subsection}{\normalsize \bfseries}
\titlelabel{\thetitle.\,\,}

\usepackage[unicode=true,
bookmarks=true,
colorlinks=true,
citecolor=blue,
breaklinks=true,
letterpaper]{hyperref} 

\usepackage[backend=biber,
style=nature, 
maxbibnames=99,
maxcitenames=2,
mincitenames=2,
uniquename=false,
isbn=false, 
doi=false,
url=true
]{biblatex}
\renewbibmacro{in:}{%
  \ifentrytype{article}{}{\printtext{\bibstring{in}\intitlepunct}}}
\usepackage{xpatch}
\xpatchbibdriver{article}
  {\usebibmacro{title}%
   \newunit}
  {\usebibmacro{title}%
   \printunit{\addcomma\space}}
  {}
  {}
\renewbibmacro*{volume+number+eid}{%
  \printfield{volume}%
  \setunit*{\addnbspace}% NEW (optional); there's also \addnbthinspace
  \printfield{number}%
  \setunit{\addcomma\space}%
  \printfield{eid}}
\DeclareFieldFormat[article]{number}{\mkbibparens{#1}}

\DefineBibliographyStrings{english}{%
  andothers = {\textit{et al}\adddot}
}

\DeclareNameAlias{sortname}{first-last}
\DeclareFieldFormat{pages}{#1}

\bibliography{main}
\renewcommand{\cite}[1]{\parencite{#1}}

\usepackage{setspace}

\usepackage{tgpagella} % for font
\usepackage{newtxmath} % increase math font weight
\usepackage{microtype} % increase math font weight

\title{A Causal Machine Learning Framework for Predicting Preventable Hospital Readmissions}

\usepackage{authblk}
\author[1]{Ben J. Marafino}
\author[2]{Alejandro Schuler}
\author[2]{Vincent X. Liu}
\author[2]{Gabriel J. Escobar}
\author[1]{Mike Baiocchi}
\affil[1]{Stanford University}
\affil[2]{Division of Research, Kaiser Permanente Northern California}

\newdateformat{monthyeardate}{%
  \monthname[\THEMONTH], \THEYEAR.}
\date{Current version: \monthyeardate\today}

\begin{document}
\begin{spacing}{2.0}

\begin{spacing}{1.0}
\maketitle
\vspace{-4ex}
\begin{abstract}
Clinical predictive algorithms are increasingly being used to form the basis for optimal treatment policies---that is, to enable interventions to be targeted to the patients who will presumably benefit most. Despite taking advantage of recent advances in supervised machine learning, these algorithms remain, in a sense, blunt instruments---often being developed and deployed without a full accounting of the causal aspects of the prediction problems they are intended to solve. Indeed, in many settings, including among patients at risk of readmission, the riskiest patients may derive less benefit from a preventative intervention compared to those at lower risk. Moreover, targeting an intervention to a population, rather than limiting it to a small group of high-risk patients, may lead to far greater overall utility if the patients with the most modifiable (or preventable) outcomes across the population could be identified. Based on these insights, we introduce a causal machine learning framework that decouples this prediction problem into causal and predictive parts, which clearly delineates the complementary roles of causal inference and prediction in this problem. We estimate treatment effects using causal forests, and characterize treatment effect heterogeneity across levels of predicted risk using these estimates. Furthermore, we show how these effect estimates could be used in concert with the modeled "payoffs" associated with successful prevention of individual readmissions to maximize overall utility. Based on data taken from before and after the implementation of a readmissions prevention intervention at Kaiser Permanente Northern California, our results suggest that nearly four times as many readmissions could be prevented annually with this approach compared to targeting this intervention using predicted risk.
\end{abstract}
\end{spacing}

\section{Introduction: Hospital Readmissions and the Pitfalls of Risk-Based Targeting}

Unplanned hospital readmissions represent an undesirable outcome following a hospitalization, but are common, costly, and associated with substantial morbidity and mortality, occurring within 30 days following nearly 20\% of hospitalizations by Medicare beneficiaries \cite{Jencks2009RehospitalizationsProgram}. In 2011, 3.3 million patients in the United States were readmitted to the hospital within 30 days, incurring costs of \$41 billion \cite{Hines2011Conditions2011}. In 2012, responding the growing awareness of the toll of readmissions, the Centers for Medicare and Medicaid Services introduced the Hospital Readmissions Reduction Program (HRRP), which penalizes hospitals with risk-adjusted 30-day readmission rates higher than the average. As a consequence of the HRRP and other value-based care initiatives, many hospitals and health care systems in the United States have since implemented quality improvement (QI) initiatives and population health management programs that rely on risk assessment tools to identify hospitalized patients at high risk of readmission. Tailored interventions can be then targeted to these patients immediately following discharge, with the goal of preventing their readmission. The effectiveness of these interventions in preventing readmissions has been mixed, and the precise mechanisms through which they do so remain unclear. \cite{Leppin2014PreventingTrials,Hansen2011InterventionsReview,Wadhera2018AssociationPneumonia,Kansagara2016SoLiterature,Finkelstein2020HealthTrial,Bates2014BigPatients,Berkowitz2018AssociationJ-CHiP}

Many risk assessment tools used in these efforts apply statistical modeling or supervised machine learning to estimate readmission risk among hospitalized patients based on data prior to discharge.  \cite{Kansagara2011RiskReview.,Bayati2014Data-drivenStudy.,Escobar2015NonelectiveMortality,Billings2006CasePatients,Berkowitz2018AssociationJ-CHiP,Bates2014BigPatients} Stakeholders select a risk threshold with respect to resource constraints, which underpins objective criteria (often referred to in the machine learning literature as a \textit{treatment policy}) that determine which patients are selected to receive the intervention. Common threshold-based criteria specify that an intervention is to be delivered to all patients above a prespecified risk threshold, while those below it receive usual care. Underlying many population health management and QI efforts aimed at reducing readmissions is the implicit assumption that the patients most at risk are also those most likely to benefit from the intervention. \cite{Fihn2014InsightsAdministration,Bates2014BigPatients,HealthITAnalytics2016UsingHealthitanalytics.com/features/using-risk-scores-stratification-for-population-health-management,HealthITAnalytics2018TopHttps://healthitanalytics.com/news/top-4-big-data-analytics-strategies-to-reduce-hospital-readmissions} Ostensibly, this assumption has intuitive appeal, given that higher-risk patients appear to have "more room to move the needle", but it is not guaranteed to hold in practice \cite{Ascarza2018RetentionIneffective, Athey2017BeyondProblems}, especially in the context of readmissions \cite{Finkelstein2020HealthTrial,Lindquist2011UnderstandingFactors} and other settings where treatment effect heterogeneity may be present \cite{Athey2017BeyondProblems}.

The need for analytical approaches that estimate patient-level benefit---referred to in some contexts as \textit{impactibility} \cite{Lewis2010ImpactibilityPrograms,Freund2011IdentificationPrograms,Steventon2017PreventingRisk,Flaks-Manov2020PreventingPrediction}---is beginning to be recognized, particularly for readmission reduction programs \cite{Steventon2017PreventingRisk}. However, the distinction between benefit and risk does not yet appear to be widely appreciated by both those developing and applying risk assessment tools. Individual benefit is often expressed in terms of treatment effects, which cannot be estimated by modelling outcome risk. Predicting, for example, a readmission risk of 60\% for a patient provides no information on their counterfactual risk if they were to receive a readmissions reduction intervention. The actual counterfactual risk for this hypothetical patient could be unchanged, on average, corresponding to no effect for the intervention. On the other hand, the effect of this intervention may be heterogeneous across levels of predicted risk, so that, for example, this patient experiences an absolute risk reduction (ARR) of 10\% as a result of the intervention, while another patient at a predicted risk of 30\% experiences an ARR of 20\%. Given limited resources, a decision-maker may wish to give the intervention to the latter patient. Indeed, when it comes to preventing readmissions, there is growing evidence that higher-risk patients---referred to in some contexts as "super-utilizers" \cite{Finkelstein2020HealthTrial}---may be less sensitive to a class of care coordination interventions relative to those at lower risk \cite{Steventon2017PreventingRisk,Lindquist2011UnderstandingFactors,Rich1993PreventionStudy.}. 

Moreover, efforts targeting preventative interventions based on predicted risk also fail to take into account that low-risk patients comprise the majority of readmissions \cite{Roland2012ReducingTrack}. That the majority of poor outcomes are experienced by patients at low risk, but who would not have been selected to receive an intervention, is an observation which has also surfaced in a range of predictive modeling problems in population health management \cite{Bates2014BigPatients}. Thus, targeting preventative interventions so as to include lower-risk patients among whom they may be effective, rather than targeting them only to high-risk patients, may potentially prevent more readmissions than the latter strategy \cite{Rose1985SickPopulations,Chiolero2015TheStrategy, McWilliams2017FocusingCosts}. However, even given a hypothetical ideal intervention---one guaranteed to be effective for \textit{all} patients in a population---staffing and other resource constraints may preclude scaling up such an intervention to an entire population. Hence, in order to maximize the net benefit of an intervention, "decoupling" the prediction problem into causal and predictive components---modeling not just heterogeneity in treatment effects, but also the (possibly heterogeneous) "payoffs" associated with successful prevention---may be necessary \cite{Kleinberg2015PredictionProblems}. 

Few, if any, analytical approaches to identify "care-sensitive" patients, or those whose outcomes may be most "impactible", currently exist, despite the clear need for such approaches. \cite{Lewis2010ImpactibilityPrograms,Flaks-Manov2020PreventingPrediction} Existing approaches based on off-the-shelf supervised machine learning methods, despite their flexibility and potential predictive power, cannot meet this need. \cite{Athey2017BeyondProblems} In this paper, we propose and demonstrate the feasibility of a causal machine learning framework to identify preventable hospital readmissions with respect to a readmission prevention intervention. In our context, the "preventability" of a readmission is not based on predefined, qualitative criteria, as in prior work (e.g., \cite{Goldfield2008IdentifyingReadmissions,Auerbach2016PreventabilityPatients}). Rather, it is expressed in quantitative terms: the greater the treatment effect on readmission estimated for a patient, the more preventable their potential readmission may be.

To do so, we leverage a rich set of data drawn from before and after the roll-out of a comprehensive readmissions prevention intervention in an integrated health system, seeking to: (1) estimate the heterogeneity in the treatment effect of this intervention; (2) characterize the potential extent of mismatch between treatment effects and predicted risk; and (3) quantify the potential gains afforded by targeting based on treatment effects or benefit instead of risk. Finally, based on our findings, we also outline some possible directions for how population health management programs could be redesigned so as to maximize the aggregate benefit, or overall impact, of these preventative interventions. 

\section{Methods}

\subsection{Data and Context}

The data consist of 1,584,902 hospitalizations taking place at the 21 hospitals in Kaiser Permanente's Northern California region (hereafter KPNC) between June 2010 and December 2018. They include patient demographics, diagnosis codes, laboratory-based severity of illness scores at admission and at discharge, and a comorbidity burden score that is updated monthly. The data also record whether a patient experienced a non-elective re-admission and/or death within 30 days. These data are described in greater detail in \cite{Escobar2019MultiyearSystem}.

These data encompass a period where a comprehensive readmissions prevention intervention, known as the \textit{Transitions Program}, began and completed implementation at all 21 KPNC hospitals from January 2016 to May 2017. The Transitions Program had two goals: (1) to standardize post-discharge care by consolidating a range of preexisting care coordination programs for patients with complex care needs; and (2) to improve the efficiency of this standardized intervention by targeting it to the patients at highest risk of the composite outcome of post-discharge re-admission and/or death. 

As currently implemented, the Transitions Program relies on a validated predictive model for the risk of this composite outcome \cite{Escobar2015NonelectiveMortality}, which was developed using historical data from between June 2010 and December 2013, before the implementation of the Transitions Program. Following development and validation of this model by teams at KPNC's Division of Research, it was subsequently integrated into KP HealthConnect, KPNC's electronic health record (EHR) system to produce continuous risk scores, ranging from 0 to 100\%, at 6:00 AM on the planned discharge day. These risk scores are used to automatically assign inpatients awaiting discharge to be followed by the Transitions Program over the 30-day period post-discharge. Inpatients with a predicted risk of $\geq\!\!25$ are assigned to be followed by the Transitions Program, and are considered to have received the Transitions Program intervention. On the other hand, inpatients with a predicted risk below 25\% receive usual post-discharge care at the discretion of the discharging physician. 

The Transitions Program intervention consists of a bundle of discrete interventions aimed at improving the transition from inpatient care to home or a to a skilled nursing facility. Together, they comprise an interlocking care pathway over the 30 days following discharge, beginning on the morning of the planned discharge day, and which we summarize step-by-step here and in Table \ref{tab:transitions}.

\begin{table}[]
\centering
\resizebox{\textwidth}{!}{%
\begin{tabular}{@{}cccccc@{}}
\toprule
Risk Level & Initial Assessment & Week 1 & Week 2 & Week 3 & Week 4 \\ \midrule
\begin{tabular}[c]{@{}c@{}}High\\ ($\geq 45$\%)\end{tabular} & \begin{tabular}[c]{@{}c@{}}Phone follow-up \\ within 24 to 48 hours \\ \\ \textit{and}\end{tabular} & \begin{tabular}[c]{@{}c@{}}Phone follow-up\\ every other day\end{tabular} & \begin{tabular}[c]{@{}c@{}}2 phone follow-ups; \\ more as needed\end{tabular} & \begin{tabular}[c]{@{}c@{}}Phone follow-up \\ once weekly; \\ more as needed\end{tabular} & \begin{tabular}[c]{@{}c@{}}Phone follow-up \\ once weekly; \\ more as needed\end{tabular} \\ \cmidrule(r){1-1} \cmidrule(l){3-6} 
\begin{tabular}[c]{@{}c@{}}Medium\\ ($25-45$\%)\end{tabular} & \begin{tabular}[c]{@{}c@{}}Primary care physician follow-up visit\\ within 2 to 5 days\end{tabular} & \multicolumn{4}{c}{\begin{tabular}[c]{@{}c@{}}Once weekly phone follow-up \\ (with more as needed)\end{tabular}} \\ \midrule
\begin{tabular}[c]{@{}c@{}}Low \\ ($\leq 25$\%)\end{tabular} & \multicolumn{5}{c}{Usual care at discretion of discharging physician} \\ \bottomrule
\end{tabular}%
}
\caption{\footnotesize The Transitions Program intervention pathway. The initial assessment applies to both the medium and high risk groups. Following it, the pathway diverges in terms of the frequency of phone contact.}
\label{tab:transitions}
\end{table}

For inpatients awaiting discharge and who have been assigned to be followed by the Transitions Program, on their planned discharge day, a case manager meets with them at the bedside to provide information on the Transitions Program. Next, once the patient has arrived home, a Transitions case manager calls them within 24 to 48 hours to walk them through their discharge instructions, and to identify any gaps in their understanding of them. If necessary, the case manager can also refer the patient to a pharmacist or social worker for focused follow-up. At the same time, the nurse also works to make an appointment with the patient's primary care physician to take place within 3 to 5 days post-discharge.

Following this initial outreach, the Transitions case manager continues to contact the patient weekly by phone, and remains available throughout if the patient requires further assistance. At 30 days post-discharge, the patient is considered to have "graduated" and is no longer followed by the Transitions Program. All steps of this process are initiated through and documented in the EHR, enabling consistent followup for the patients enrolled. A special category of patients are considered at very high risk if their predicted risk is $\geq \!\! 45$\% or if they lack social support, and receive a more intensified version of the intervention. This version entails  follow-up every other day via telephone for the first week post-discharge, followed by $\geq \!\! 2$ times a week the second week, and once a week afterward until "graduation" at 30 days.

Initial analyses of the impacts of the Transitions Program \cite{Marafino2020ASystem}, using a hybrid difference-in-differences analysis/regression discontinuity approach \cite{Walkey2020NovelInitiative}, indicated that it was effective, being associated with approximately 1,200 and 300 fewer annual readmissions and deaths, respectively, within 30 days following their index discharge. Notably, these analyses also suggested some extent of risk-based treatment effect heterogeneity, in that the intervention appeared to be somewhat less effective for patients at higher risk compared to those at relatively lower risk. 

As a consequence of the conclusions of these analyses, two questions presented themselves. The first was whether the Transitions Program intervention could be re-targeted more efficiently, leading to greater aggregate benefit, possibly by including patients at lower risk who would not otherwise have received the intervention. The second arose out of the risk-based treatment effect heterogeneity suggested by these analyses---may the intervention in fact be less effective for patients at higher risk? If so, what is the extent of the mismatch between risk and treatment effects? To answer these questions requires modeling individual heterogeneous treatment effects, and not risk, as is commonly done.

In this paper, we use the same data on 1,584,902 patients as that used for the analysis of the effectiveness of the Transitions Program. As in that analysis, we use a subset of 1,539,285 "index stays" which meet a set of eligibility criteria. These criteria include: the patient was discharged alive from the hospital; age $\geq \!\! 18$ years at admission; and their admission was not for childbirth (although post-delivery complications were included) nor for same-day surgery. Moreover, we consider a readmission non-elective if it began in the emergency department; if the principal diagnosis was an ambulatory care-sensitive condition \cite{AgencyforHealthcareResearchandQuality2001AHRQConditions.}; or if the episode of care began in an outpatient clinic, and the patient had elevated severity of illness, based on a mortality risk of $\geq\!\!7.2$\% as predicted by their laboratory-based acuity score (\texttt{LAPS2}) alone. The covariates used in this study (as input to the causal forest below) are summarized in Table \ref{tab:table1}.

This project was approved by the KPNC Institutional Review Board for the Protection of Human Subjects, which has jurisdiction over all the study hospitals and waived the requirement for individual informed consent. 

\begin{table}
  \centering
\resizebox{\textwidth}{!}{%
 \begin{tabular}{ll}
    \toprule
    \texttt{AGE} & Patient age in years, recorded at admission \\
    \texttt{MALE} & Male gender indicator \\
    \texttt{DCO\_4} & Code status at discharge (4 categories) \\
    \texttt{HOSP\_PRIOR7\_CT} & Count of hospitalizations in the last 7 days prior to the current admission \\
    \texttt{HOSP\_PRIOR8\_30\_CT} & Count of hospitalizations in the last 8 to 30 days prior to the current admission \\
    \texttt{LOS\_30} & Length of stay, in days (with stays above 30 days truncated at 30 days) \\
    \texttt{MEDICARE} & Indicator for Medicare Advantage status \\
    \texttt{DISCHDISP} & Discharge disposition (home, skilled nursing, home health; one-hot encoded) \\
    \texttt{LAPS2} & Laboratory-based acuity of illness score, recorded at admission \\
    \texttt{LAPS2DC} & Laboratory-based acuity of illness score, recorded at discharge\\
    \texttt{COPS2} & Comorbidity and chronic condition score, updated monthly\\
    \texttt{HCUPSGDC} & Diagnosis super-group classification (30 groups; one-hot encoded) \\
    \midrule
    \texttt{W} (or $W_i$) & Treatment: Transitions Program intervention \\
    \texttt{Y} (or $Y_i$) & Outcome: Non-elective readmission within 30 days post-discharge \\
    \bottomrule
  \end{tabular}}
  \vspace{0.2em}
 \caption{\footnotesize A list of the covariates used in this study.}
   \label{tab:table1}
\end{table}

\subsection{From (observational) data to predicted treatment effects: Causal forests}

To identify potentially preventable readmissions, we undertake a causal machine learning approach using data taken from before and after the implementation of the Transitions Program at KPNC. Our causal machine learning approach is distinct from supervised machine learning as it is commonly applied in that it seeks to estimate individual \textit{treatment effects}, and not outcome risk. This objective cannot be readily accomplished with off-the-shelf supervised machine learning methods. \cite{Athey2017BeyondProblems} Compared to other methods for studying treatment effect heterogeneity (e.g. subgroup analyses), causal machine learning methods afford two advantages: one, they avoid strong modeling assumptions, allowing a data-driven approach; and two, they guard against overfitting through applying a form of regularization.

We express these individual treatment effects of the Transitions Program intervention in terms of the estimated conditional average treatment effects (CATE), $\hat{\tau}_i$, which can be interpreted as absolute risk reductions (ARRs). It is through the sign and magnitude of these estimated CATEs that we consider a readmission potentially preventable: a $\hat{\tau}_i < 0$ denotes that the intervention would be expected to lower 30-day readmission risk for that patient, while a $\hat{\tau}_i > 0$ suggests that the intervention would be more likely to result in readmission within 30 days. A larger (more negative) CATE suggests a greater extent of preventability: i.e., $\hat{\tau}_j < \hat{\tau}_i < 0$ implies that patient $j$'s readmission is more "preventable"---their risk is more modifiable by the intervention---compared to patient $i$'s.

To estimate these CATEs, we apply causal forests to the KPNC data described in the previous subsection. Causal forests \cite{AtheyGRF, Wager2018EstimationForests} represent a special case of generalized random forests \cite{Athey2019GeneralizedForests}. They have been used in a variety of applications, including to estimate conditional average partial effects; our overall approach resembles that undertaken in \cite{Athey}, which used them to study treatment effect heterogeneity in an observational setting. Causal forests can be viewed as a form of adaptive, data-driven subgroup analysis, and can be applied in either observational or randomized settings, relying on much of the machinery of random forests \cite{Breiman2001RandomForests}.

These machinery retained from random forests include recursive partitioning, subsampling, and random selection of the splits. However, causal forests differ from random forests in several significant ways. For one, the splitting criterion seeks to place splits so as to maximize heterogeneity, instead of minimizing prediction error as with random forests. Moreover, when placing splits, the causal forest algorithm (depending on implementation details) has access to the data $X_i$ and treatment assignments $W_i$, but not to the outcomes $Y_i$, and a separate sample is used to estimate effects once these splits have been placed---a property often referred to as 'honesty' \cite{Wager2018EstimationForests}.

We describe some necessary assumptions that are required in order to identify the CATEs, as the data are considered observational, and not randomized. The causal forest already fits a treatment assignment model, which deconfounds the subgroups in our study (described below) to some extent, but we also make further assumptions regarding the data, which we describe below, to facilitate deconfounding. We begin with some notation: for each of a set of units $i = 1, \ldots, n$, we observe the triple $(X_i, Y_i, W_i)$, where $X_i \in \mathbb{R}^p$ is a covariate vector, $Y_i \in \{0,1\}$ denotes the observed outcome, and $W_i \in \{0,1\}$ treatment assignment. Following Rubin's potential outcomes framework \cite{Rubin1974EstimatingStudies}, we assume the existence of potential outcomes, $Y_i(1)$ and $Y_i(0)$, for each unit, and define the conditional average treatment effect (CATE) for an unit $i$ with the covariate vector $X_i = x$ as 

\begin{equation}
    \tau_i(x) = \mathbb{E}[Y_i(1) - Y_i(0) \mid X_i = x]. 
\end{equation}

Within a leaf $L$, a causal forest estimates this quantity as 
\begin{equation}
    \hat{\tau}(x) = \frac{1}{\mid\!\! \{j : W_j = 1 \wedge X_j \in L \}\!\! \mid} \sum_{\{j : W_j = 1 \wedge X_j \in L\}} Y_j \quad - \quad \frac{1}{\mid\!\! \{ j : W_j = 0 \wedge X_j \in L \}\!\! \mid} \sum_{\{j : W_j = 0 \wedge X_j \in L\}} Y_j
\end{equation}

for a $x \in L$. Heuristically, the process of fitting a causal forest aims to make these leaves $L$ as small as possible so that the data in each resemble a randomized experiment, while simultaneously maximizing effect heterogeneity. \cite{Wager2018EstimationForests}

However, as we observe only one of the two potential outcomes for each unit, $Y_i = Y_i(W_i)$, we cannot estimate $Y_i(1) - Y_i(0)$ directly from these data. Under some assumptions, however, we can reprovision the units in the data that did experience the counterfactual outcome to estimate $\tau_i$, by having those units serve as 'virtual twins' for an unit $i$. These assumptions entail (1) the existence of these twins; and (2) that, in some sense, these twins look similar in terms of their covariates. These are the \textit{overlap} and \textit{uncounfoundedness} assumptions, respectively. Heuristically, the overlap assumption presumes that these twins could exist, and unconfoundedness posits that these twins are in fact similar in terms of their observed covariates. Together, these assumptions allow us to have some confidence that the $\tau_i(x)$ in fact do identify causal effects.

We address identification with respect to the predicted risk threshold of 25\%, as well as the time period. By time period, recall that the Transitions Program intervention was rolled out to each of the 21 KPNC hospitals in a way that formed two completely disjoint subsets of patients, corresponding to the pre- and post-implementation periods. Also recall that patients were assigned to the intervention if they were discharged during the post-implementation period and their risk score was $>\!\!25$\%, while those below that value received usual care. An useful feature of these data is that all patients in the pre-implementation period were assigned 'shadow' risk scores using the same instantiation of the predictive model, even though none of these patients received treatment. Hence, the data are split into four disjoint subgroups, indexed by risk category and time period: 
\[
(\text{pre}, \geq\!\!25), \quad (\text{post}, \geq\!\!25), \quad (\text{pre}, <\!\!25), \quad (\text{post}, <\!\!25), 
\]

i.e., the tuple $(\text{pre}, \geq\!\!25)$ denotes the subgroup consisting of patients discharged during the pre-implementation period with a risk score of $\geq\!\!25$\%, and $(.\,, \geq\!\!25)$ denotes all patients with risk $\geq\!\!25$\% in the data. Each hospital discharge belongs to one and only one of these subgroups.

Heuristically, these 'shadow' risk scores allow us to mix data from across periods so that the $(\text{pre}, \geq\!\!25)$ subgroup can be used as a source of counterfactuals for patients in the $(\text{post}, \geq\!\!25)$ subgroup, assuming no unmeasured confounding as a consequence of the shift from "pre" to "post". Stratifying on the risk score allows us to deconfound the potential outcomes of the patients in these two subgroups. Moreover, these two subgroups together can be used to provide plausible counterfactuals for the patients in the $(.\,, <\!\!25)$ risk subgroup, despite none of those patients having been assigned to the intervention. We describe the identification strategy---which relies on standard ignorability assumptions---in more detail below, beginning with the $(.\,, \geq\!\!25)$ subgroup.

First, for each of the four subgroups, we assume overlap: given some $\epsilon > 0$ and all possible $x \in \mathbb{R}^p$, 

\begin{equation}
    \epsilon < P(W_i = 1 \mid X_i = x) < 1 - \epsilon.
\end{equation}

This assumption means that no patient is guaranteed to receive the intervention, nor are they guaranteed to receive the control, based on their covariates $x$. 

For the patients in the $(.\,, \geq\!\!25)$ group, our identification strategy makes use of the balancing properties of risk scores (or prognostic scores) \cite{Hansen2008TheScore} to establish unconfoundedness.  Assuming no hidden bias, conditioning on a prognostic score $\Psi(x) = P(Y \mid W = 0,  x)$ is sufficient to deconfound the potential outcomes. The risk score used to assign the Transitions intervention is a prognostic score; hence, it is sufficient to deconfound these potential outcomes.

For patients in the $(.\,, <\!\!25)$ subgroup, the picture is slightly more complicated. Among these patients, we cannot assume exchangeability conditional on the risk score $\hat{Y}_i$, 

\begin{equation}
   \{ Y_i(1), Y_i(0) \} \bigCI W_i \mid \hat{Y}_i,
\end{equation}

because, again, treatment assignment is contingent on a predicted risk $\hat{Y}_i \geq 0.25$, i.e., $W_i = \mathbf{1}\{\hat{Y}_i \geq 0.25\}$. However, recall that the causal forests are performing estimation in $X_i$-space, and not in $\hat{Y}_i$-space, and note that we can instead impose the slightly weaker assumption of ignorability conditional on some subset $X'_i \subseteq X_i$, which comprise inputs to the score $\hat{Y}_i$; namely, that 

\begin{equation}
   \{ Y_i(1), Y_i(0) \} \bigCI W_i \mid X'_i,
\end{equation} 

which we can justify \textit{a priori} with the knowledge that no one component predictor predominates in the risk model (see the appendix to \cite{Escobar2015NonelectiveMortality}); that is, no one covariate strongly determines treatment assignment. We provide empirical evidence to establish the plausibility of this assumption, at least in low dimensions, in Figure \ref{fig:overlap}. Together with assuming the existence of potential outcomes, this \textit{unconfoundedness} assumption is sufficient to obtain consistent estimates of $\tau(x)$ \cite{Wager2018EstimationForests}. Moreover, since causal forests perform a form of local estimation, our assumptions are independent for the $(.\,, \geq\!\!25)$ and $(.\,, <\!\!25)$ subgroups in the sense that if the unconfoundedness assumption fails for either subgroup, but not the other, the estimates for the subgroup in which it does hold should not be affected.

Finally, as a formal assessment of treatment effect heterogeneity, we also perform the omnibus test for heterogeneity \cite{Chernozhukov2017}, which seeks to estimate the best linear predictor of the CATE by using the "out-of-bag" predictions from the causal forest, $\hat{\tau}^{-i}$, to fit the following linear model:

\begin{equation}
    Y_i - \hat{m}^{-i}(X_i) = \alpha\bar{\tau}(W_i - \hat{e}^{-i}(X_i)) + \beta(\hat{\tau}^{-i}(X_i) - \bar{\tau})(W_i - \hat{e}^{-i}(X_i)) + \epsilon,
\end{equation}

where

\begin{equation}
    \bar{\tau} = \frac{1}{n} \sum^n_{i=1} \hat{\tau}^{-i}(X_i),
\end{equation}

and $\hat m(.)$ and $\hat e(.)$ denote the marginal outcome and assignment models estimated by the causal forest, respectively. (The superscript $-i$ denotes that the quantity was computed "out-of-bag", i.e., that the forest was not trained on example $i$). Fitting this linear model yields two coefficient estimates, $\alpha$ and $\beta$; an interpretation of these coefficients is that $\alpha$ captures the average treatment effect, and if $\alpha \approx 1$, then the predictions the forest makes are correct, on average. Likewise, $\beta$ measures how the estimated CATEs covary with the true CATEs; if $\beta \approx 1$, then these CATE estimates are well-calibrated. Moreover, we can use the $p$-value for $\beta$ as an omnibus test for heterogeneity; if the coefficient is statistically significantly greater than zero, then we can reject the null hypothesis of no treatment effect heterogeneity. \cite{AtheyEstimatingApplication}

All analyses were performed in R (version 3.6.2); causal forests and the omnibus test for heterogeneity were implemented using the \texttt{grf} package (version 0.10.4). Causal forests were fit using default settings with $n = 8,000$ trees and per-hospital clusters (for a total of 21 clusters).

\subsection{Translating predictions into treatment policies: Decoupling effects and payoffs}

A relevant question, once predictions have been made---be they of risk or of treatment effects---is how to translate them into treatment decisions. With predicted risk, these decisions are made with respect to some risk threshold, or to a decision-theoretic threshold that takes utilities into account (e.g. as in the approach in \cite{Bayati2014Data-drivenStudy.}). However, both approaches are potentially suboptimal in the presence of treatment effect heterogeneity, requiring strong assumptions to be made regarding the nature of the treatment effect. (Indeed, \cite{Bayati2014Data-drivenStudy.} assumes a constant treatment effect for all patients.) 

Here, however, we deal with treatment effects instead of risk; an obvious approach starts by treating all patients $i$ with $\hat{\tau}(X_i) < 0$---that is, by treating all patients who are expected to benefit. However, resources may be constrained so that it is infeasible to treat all these patients, and making it necessary to prioritize from among those with $\hat{\tau}_i < 0$. One way to do so is to incorporate the costs associated with the potential outcome of a readmission, or the "payoffs" $\pi$ associated with successfully preventing a readmission \cite{Kleinberg2015PredictionProblems}, which we denote by $\pi_i = \pi(X_i)$. 

There are several ways to characterize these payoffs, which ideally can be done mainly in terms of the direct costs required to provide care for a readmitted patient, as well as financial penalties associated with high readmission rates. However, these data are not available to us, so we instead use length of stay (LOS) from the readmission as a proxy for cost, and assume that higher LOS is associated with higher resource utilization and thus higher costs. A range of payoffs could be specified, such as the risk of in-hospital mortality during the readmission, or an acuity-scaled LOS measure, as well as weighted combinations of these quantities. It is important to emphasize that these payoffs are associated with the characteristics of the readmission following the index stay, if one does occur---not those of the index stay itself. 

One approach to estimating these payoffs is to predict them using historical data, i.e., $\hat{\pi}(X_i) = \mathbb{E}[\pi_i \mid X_i = x]$ in a manner similar to that used to derive the risk scores. However, this is beyond the scope of this paper, and so we make some simplifying assumptions regarding the payoffs. Namely, we assume that (1) the individual payoffs $\pi_{i}$ can be approximated by the mean payoff across all patients, $\pi_{i} \approx \mathbb{E}[\pi_{i}]$, and (2) that the payoffs are mean independent of the predicted treatment effects, $\mathbb{E}[\pi_i \mid \tau_i] = \mathbb{E}[\pi_i]$. These two assumptions make it so that the $\hat{\tau}_i$ become the sole decision criterion for the treatment policies we evaluate in this paper, but we briefly outline how to incorporate these payoffs in decision-making. 

Given both the predicted treatment effects, $\hat{\tau}_i$, and payoffs, $\hat{\pi}_i$, we can compute the individual expected utilities, $\mathbb{E}[u_i] = \hat{\tau}_i \hat{\pi}_i$ for each patient. We assume that decision-makers are risk-neutral and that the cost to intervene is fixed. Then, given two patients, $i$ and $j$, and their respective expected utilities, we would prefer to treat $i$ over $j$ if $\mathbb{E}[u_i] > \mathbb{E}[u_j]$. Another interpretation (in a population sense) is that ordering the discharges in terms of their $\hat{\tau}_i$ induces one rank ordering, while ordering them in terms of their $\mathbb{E}[u_i]$ induces another. We can treat the top $k$\% of either ordering, subject to resource constraints, but doing so with the latter will result in greater net benefit and thus would be preferred. Under the assumptions we make above, $\mathbb{E}[u_i] \propto \hat{\tau}_i$ for each patient $i$. This particular decision-theoretic approach requires absolute, and not relative outcome measures, such as the relative risk reduction. \cite{Sprenger2017ThreeMeasures}

\subsection{Measuring the impacts of different targeting strategies}

To estimate the impact (in terms of the number of readmissions prevented) of several notional targeting strategies that focus on treating those with the largest expected benefit, and not those at highest predicted risk, we undertake the following approach. We stratify the patients in the dataset into ventiles $V_1, \ldots, V_{20}$ of predicted risk, where $V_1$ denotes the lowest (0 to 5\%) risk ventile, and $V_{20}$ the highest (95 to 100\%). Then, the causal forest is trained on data through the end of 2017, and used to predict CATEs for all patients discharged in 2018. 

First, for all patients above a predicted risk of 25\%, we compute the impact of the current risk-based targeting strategy based on the predicted CATEs from 2018, and compare it to that from prior work which estimated based on the average treatment effect of this intervention. \cite{Marafino2020ASystem} This comparison serves as one check of the calibration of the predicted CATEs; the number of readmissions prevented should substantially agree in both cases. 

Second, we then use these same predicted CATEs for 2018 to assess the impact of three CATE-based targeting strategies, which treat the top 10\%, 20\%, and 50\% of patients in each risk ventile based on their predicted CATE. These strategies spread the intervention across ventiles as a form of a hedge ("not putting all one's eggs in one basket"), rather than treating the top $k$\% patients in the dataset.

The impacts of all targeting strategies are characterized both in terms of the annual number of readmissions prevented as well as the number needed to treat (NNT) to prevent one readmission. We estimate the annual number of readmissions prevented by training the causal forest on data through the end of 2017, and then summing the predicted CATEs for discharges taking place in 2018. The annual number of readmissions prevented can be expressed as

\begin{equation}
    \sum^{20}_{j = 1} \sum_{i \in V^T_j} \hat{\tau}_i,
\end{equation}

where $V^T_j$ denotes the set of patients notionally selected for treatment in ventile $j$, and $\hat{\tau}_i$ the predicted CATE (treatment effect) for the patient $i$. 

Finally, the NNT in this setting has a slightly different interpretation than the one commonly encountered in many studies. Often, the NNT is cited as a summary measure based on the overall results of a trial, e.g., the average treatment effect (ATE). In this case, an ATE estimate $\hat{\tau}$, expressed in terms of the ARR, corresponds to a NNT of $1/\overline{\hat{\tau}_i} = 1/\hat{\tau}$. However, here, we instead estimate CATEs at the individual patient level and are interested in the NNTs specific to the subgroups that receive the intervention. Hence, the NNT is a property of these subgroups and can vary across subgroups as the average predicted CATE $\overline{\hat{\tau}_i}$ varies.

\section{Results}

\subsection{Overall characteristics of the cohort}

From June 2010 to December 2018, 1,584,902 hospitalizations took place at the 21 KPNC hospitals represented in this sample. These included both inpatient hospitalizations as well as stays for observation. Further details regarding the overall cohort are presented in Table \ref{tab:a1}. Of these hospitalizations, 1,539,285 met the inclusion criteria, of which 1,127,778 (73.3\%) occurred during the pre-implementation period for the Transitions Program, and 411,507 (26.7\%) during the post-implementation period. Among these 411,507 hospitalizations taking place post-implementation, 80,424 (19.5\%) were predicted to be at high risk of 30-day post-discharge mortality or readmission; these patients were considered to have received the Transitions Program intervention following hospital discharge. 

\begin{figure}
    \centering
    \includegraphics[scale=0.67]{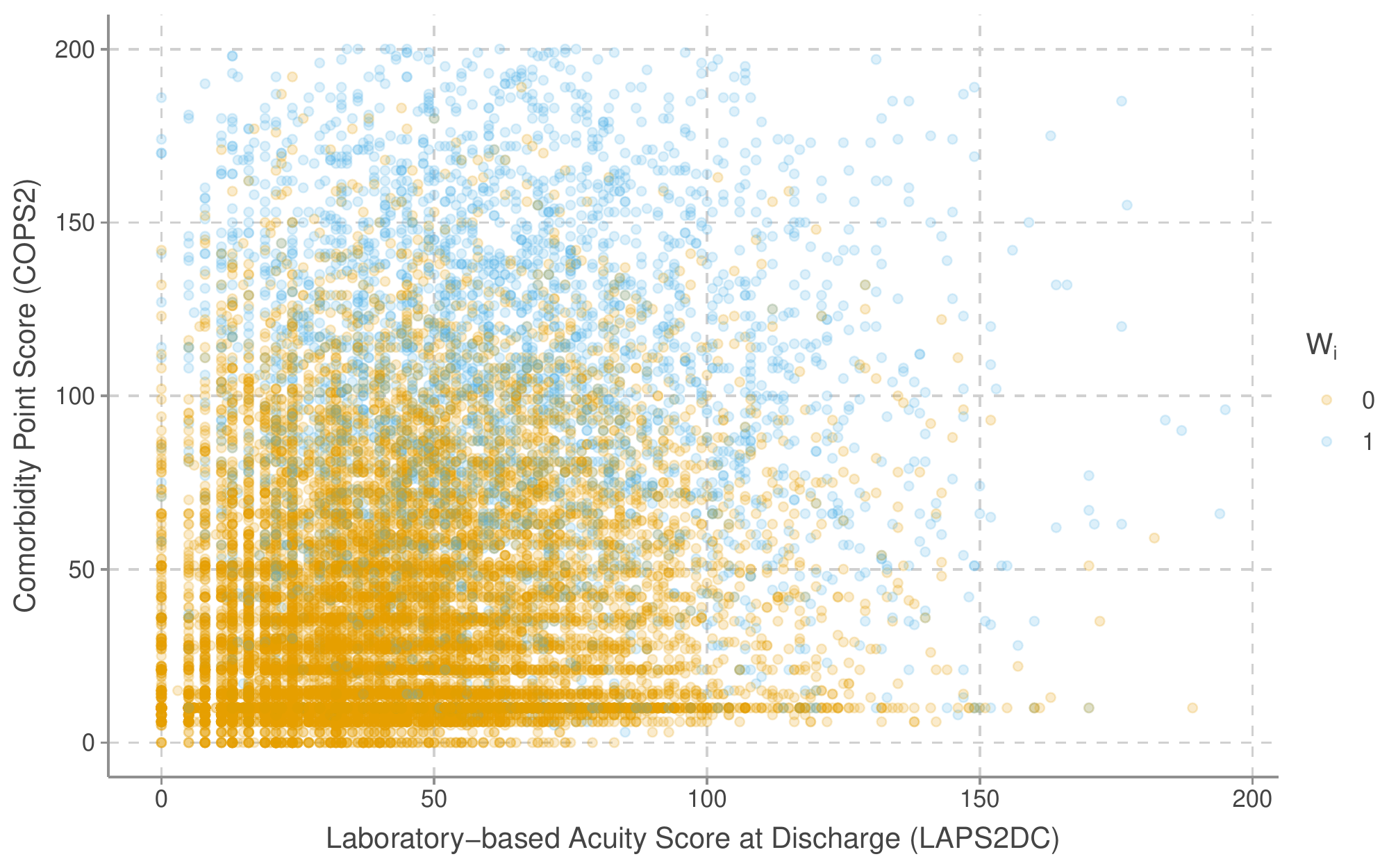}
    \caption{\footnotesize Assessing the "unconfoundedness" assumption: each point denotes an admission, which are colored according to whether they received the Transitions Program intervention post-discharge ($W_i$). The $x$-axis records the value of a laboratory-based acuity score (\texttt{LAPS2DC}) and the $y$-axis the value of a chronic condition score (\texttt{COPS2)}, both at discharge. The extent of overlap displayed here is relatively good, and implies that overlap may be implausible only among patients at very high or very low risk. This plot is based on a random sample of $n = 20,000$ index admissions taken from the post-implementation period.}
    \label{fig:overlap}
\end{figure}

Of the patients whose index stays were included, their mean age was 65.0 years, and 52.5\% were women. The overall 30-day non-elective rehospitalization rate among these index stays was 12.4\%, and 30-day post-discharge mortality was 4.0\%. Other patient-level characteristics are presented in Table \ref{tab:a1} in the Appendix. Notably, based on the distributions of \texttt{COPS2} and \texttt{LAPS2}, a key modeling assumption---that of overlap---appears to have been satisfied (Figure \ref{fig:overlap}).

\begin{figure}
    \centering
    \includegraphics[scale=0.60]{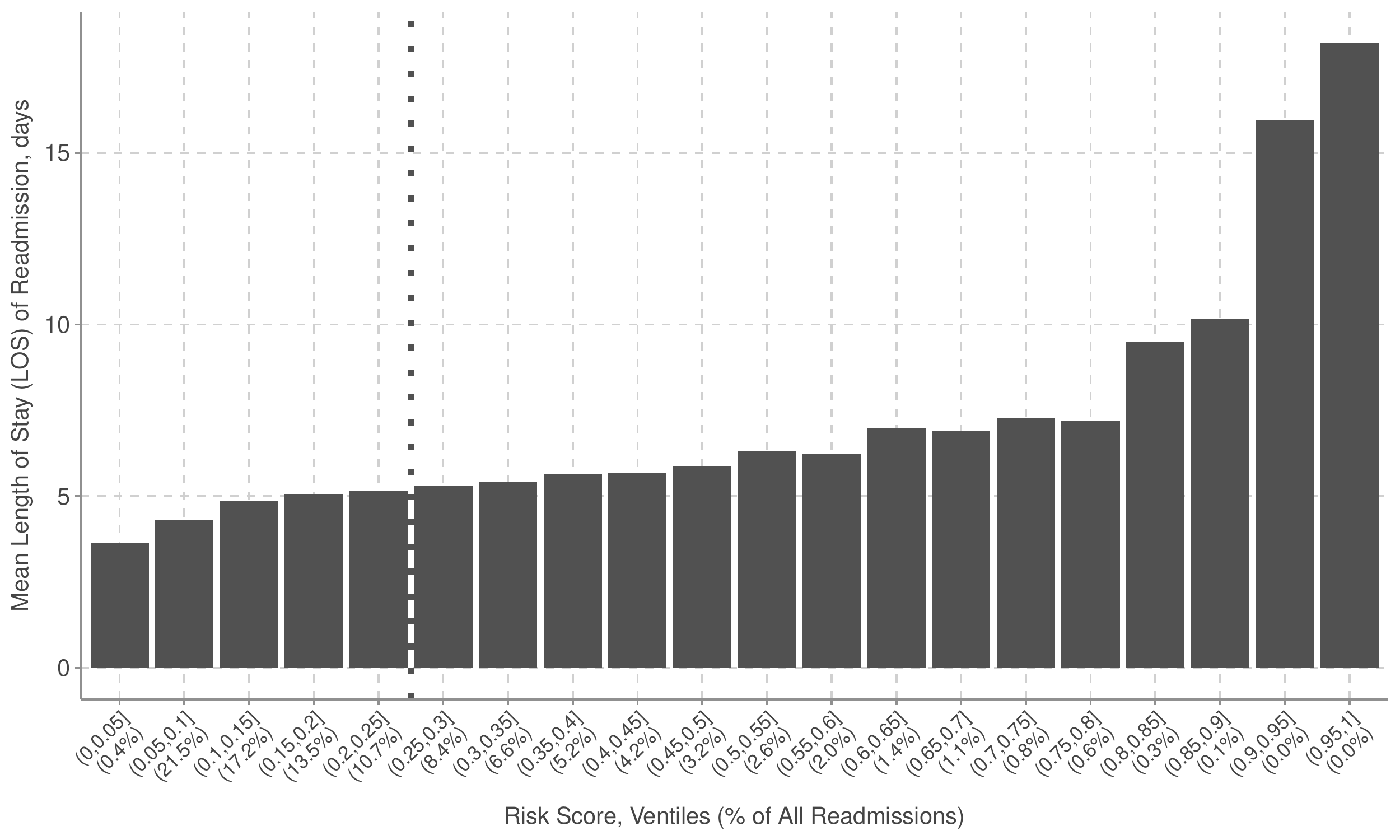}
    \caption{\footnotesize  Average length of stay (LOS) by risk score ventile. The values in parentheses below the name of each ventile denote the proportion of all 30-day readmissions incurred by patients in that ventile; patients with a predicted risk below 25\% based on their index stay accounted for 63\% of all readmissions. Notably, the average LOS is roughly similar (at 5 days) for patients with predicted risk of 5\% to 80\%. The vertical dotted line represents the 25\% risk threshold used to assign the Transitions Program intervention.}
    \label{fig:los-v-risk}
\end{figure}

Patients at low risk (risk score $<\!\!25$\%) represented 63.3\% of all readmissions throughout the study period, while making up 82.9\% of index stays, compared to 36.7\% of all readmissions among those at high risk ($\geq \!\! 25$\%), which represented 17.1\% of index stays. Moreover, the mean length of stay of the readmission following an index stay was approximately constant across ventiles of predicted risk, satisfying another assumption; patients with predicted risk of 5 to 50\% at their index discharge had a mean length of stay during their readmission that ranged from 4.6 to 5.7 days, and these patients represented 90.5\% of all readmissions. (Figure \ref{fig:los-v-risk})

\subsection{Characterizing the treatment effect heterogeneity of the Transitions Program intervention}

\begin{figure}
    \centering
    \includegraphics[scale=0.67]{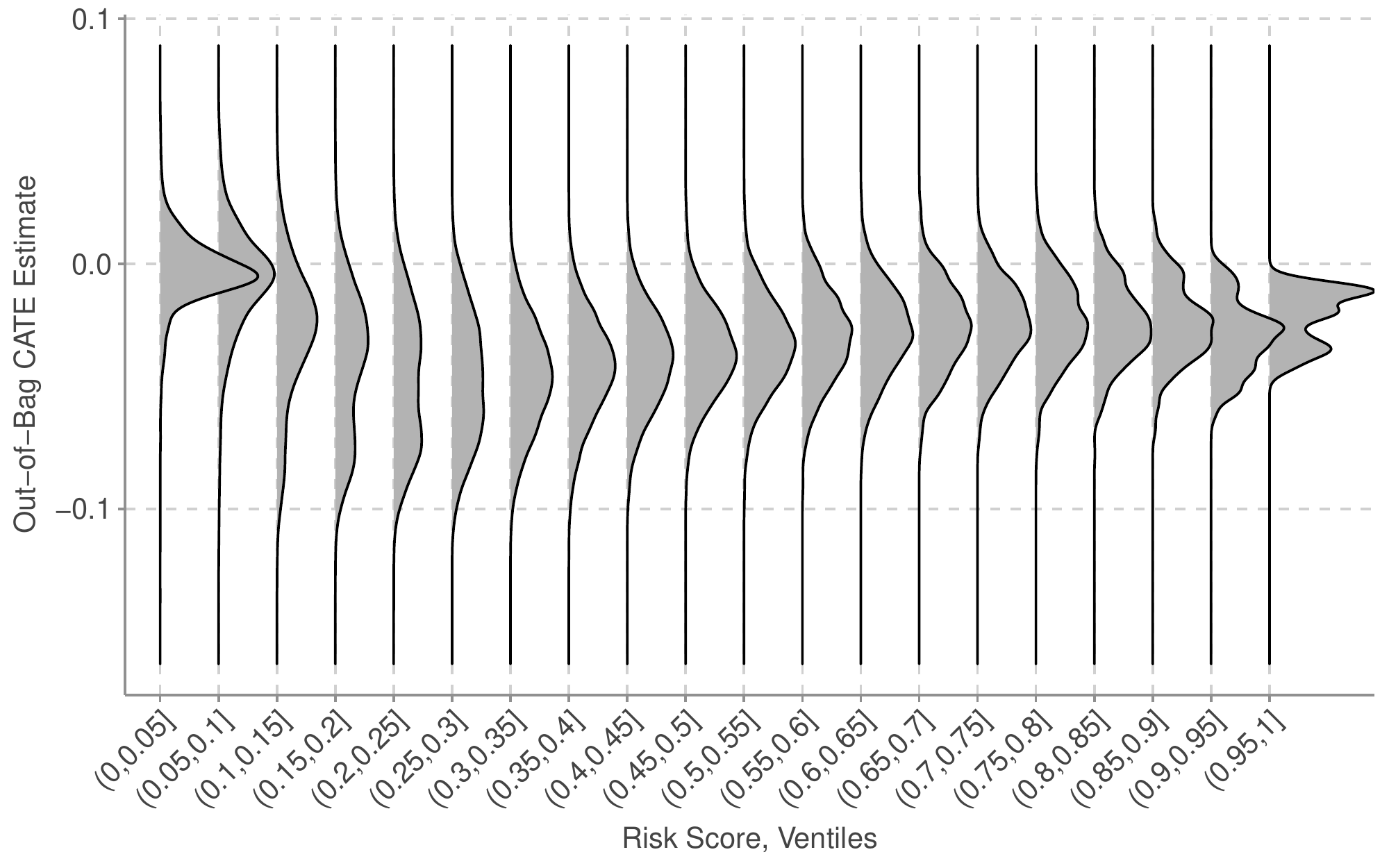}
    \caption{\footnotesize Treatment effect heterogeneity across risk score ventiles. The densities represent the distribution of estimated conditional average treatment effects within each ventile. They are drawn on a common scale, and hence do not reflect the variation in sample size across ventiles.}
    \label{fig:cate-overall}
\end{figure}

The estimated out-of-bag conditional average treatment effects (CATEs) yielded by the causal forest are presented in Figure \ref{fig:cate-overall}. The distributions of the CATEs are those of the discharges in their respective risk ventiles, but are not drawn on a common scale and so do not reflect the variation in sample size across ventiles. Qualitatively, these distributions exhibit wide spread, and suggest some extent of heterogeneity in the treatment effect of the Transitions Program intervention. In particular, treatment effects appear to be largest for patients discharged with a predicted risk of around 15 to 35\%. These effects also appeared to be somewhat attenuated as risk increased, with the center of mass tending towards zero for patients at higher risk. Notably, particularly among patients at higher risk, some estimated effects were greater than zero, indicating that the intervention was more likely to lead to readmission within 30 days. Finally, we also note that the CATE estimates themselves were well-calibrated in the sense that we identified no cases where an individual's CATE estimate was greater than their predicted risk. 

\begin{figure}
    \centering
    \includegraphics[scale=0.67]{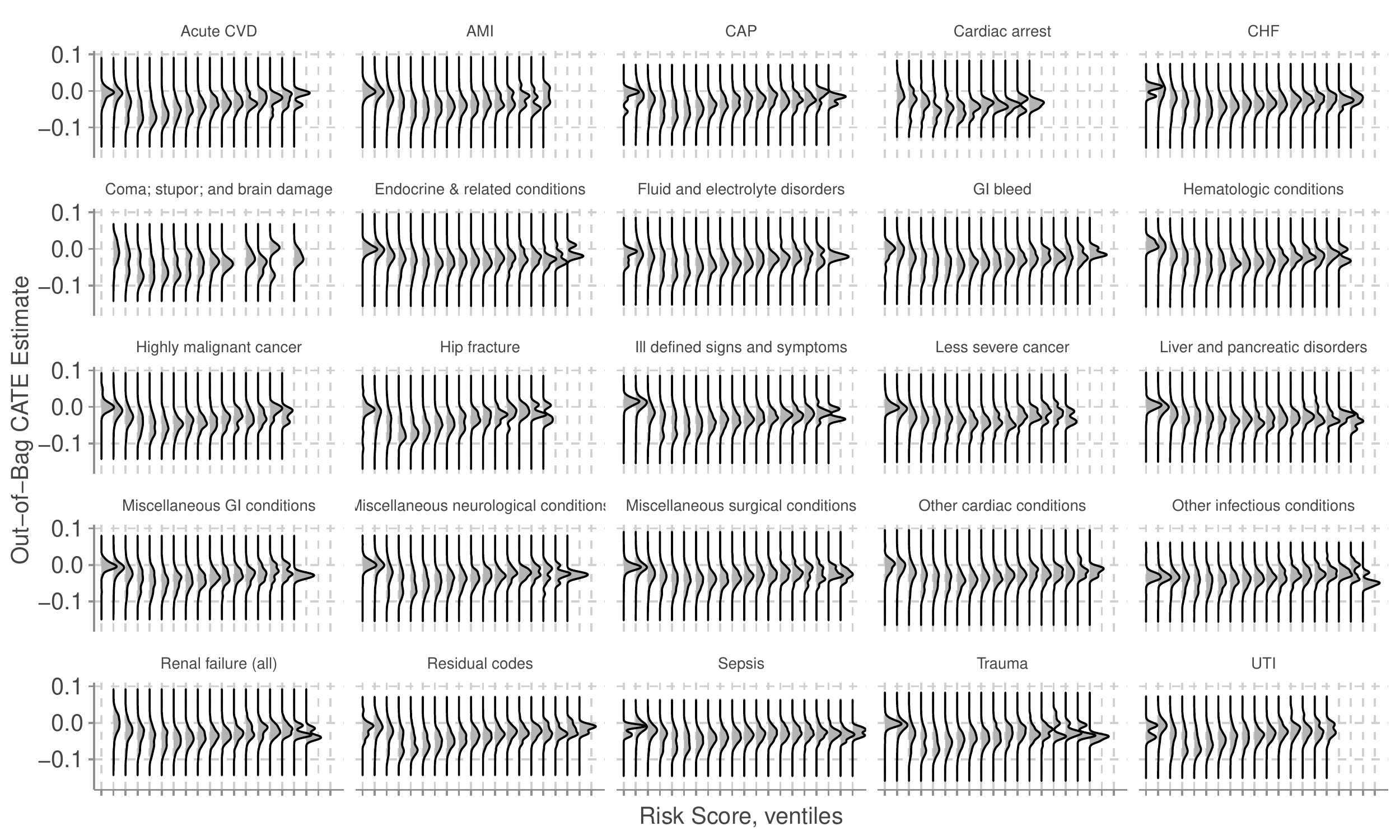}
    \caption{\footnotesize Treatment effect heterogeneity across risk score ventiles, stratified by Clinical Classification Software (CCS) supergroups based on the principal diagnosis code at discharge. A full listing of the definitions of these supergroups is given in Table \ref{tab:b1} in the Appendix. Abbreviations: CVD, cerebrovascular disease; AMI, acute myocardial infarction; CAP, community-acquired pneumonia; CHF, congestive heart failure; GI, gastrointestinal; UTI, urinary tract infection.}
    \label{fig:cate-hcupsg}
\end{figure}

Figure \ref{fig:cate-hcupsg} is similar to the previous, but stratifies the display by Clinical Classification Software (CCS) supergroups. Definitions of these supergroups can be found in Table \ref{tab:b1} in the Appendix. The overall pattern is similar to that in the unstratified plot, in that treatment effects appear to be greatest for patients at low to moderate risk, but the shapes of these distributions vary from supergroup to supergroup. All supergroups appear to exhibit heterogeneity in treatment effect within ventiles, as well, which is more pronounced for some conditions, including hip fracture, trauma, and highly malignant cancers. Qualitatively, some supergroups exhibit bimodal or even trimodal distributions in the treatment effect of the Transitions Program intervention, suggesting identification of distinct subgroups based on these effects. Some ventiles are blank for some supergroups, because there were no patients belonging to those supergroups with predicted risks falling within those ranges. 

Quantitatively, fitting the best linear predictor yields estimates of $\hat \alpha = 1.16$ and $\hat \beta = 1.06$, with $p = 5.3 \times 10^{-8}$ and $2.23 \times 10^{-7}$, respectively. Interpreting the estimate of $\beta$ as an omnibus test for the presence of heterogeneity, we can reject the null hypothesis of no treatment effect heterogeneity. 

These effects can also be evaluated on a grid of two covariates to assess how the estimated CATE function varies with interaction of these covariates. This yields insight into the qualitative aspects of the surface of the CATE function and may identify subgroups among which the Transitions Program intervention may have been more or less effective. Here, we choose the Comorbidity Point Score (\texttt{COPS2}) and the Laboratory-based Acuity Score at discharge (\texttt{LAPS2DC}), while holding all other continuous covariates at their median values, except for age, which we set to 50 and 80. Categorical covariates were held at their mode, except for the supergroup, which we set to chronic heart failure (CHF). We plot the CATE function from the 10th to 90th percentiles of \texttt{LAPS2DC} and from the 0th to 95th percentiles of \texttt{COPS2}. This is akin to evaluating the CATE function for a set of pseudo-patients with CHF having these values of \texttt{COPS2} and \texttt{LAPS2DC}. 

Figure \ref{fig:cate-hcupsg} shows the resulting CATE functions for two choices of patient age: 50 and 80 years. In this region, the estimated CATE ranged from -0.060 to 0.025 (-6.0 to 2.5\%), meaning that the estimated absolute risk reduction of the Transitions Program intervention was as large as -6\% for some patients, while for others, their readmission risk was increased by as much as 2.5\%. The estimated CATE generally was increased in magnitude---suggesting that the Transitions Program intervention became more effective as age increased---at age 80 compared to 50. Moreover, and notably, the estimated CATE tended to increase with increasing \texttt{LAPS2DC}, which measures how acutely ill a patient was upon discharge based on their laboratory test data. This finding suggests that, for patients who were more ill at discharge (indeed, the average \texttt{LAPS2DC} in 2018 was 45.5), enrolling them in the Transitions Program may actually have encouraged them to return to the hospital. While this finding of such an effect may appear surprising, it is unclear if it actually represents "harm" in the sense it is usually interpreted; we discuss this finding in more depth in the Discussion section.  

\begin{figure}
    \centering
    \includegraphics[scale=0.67]{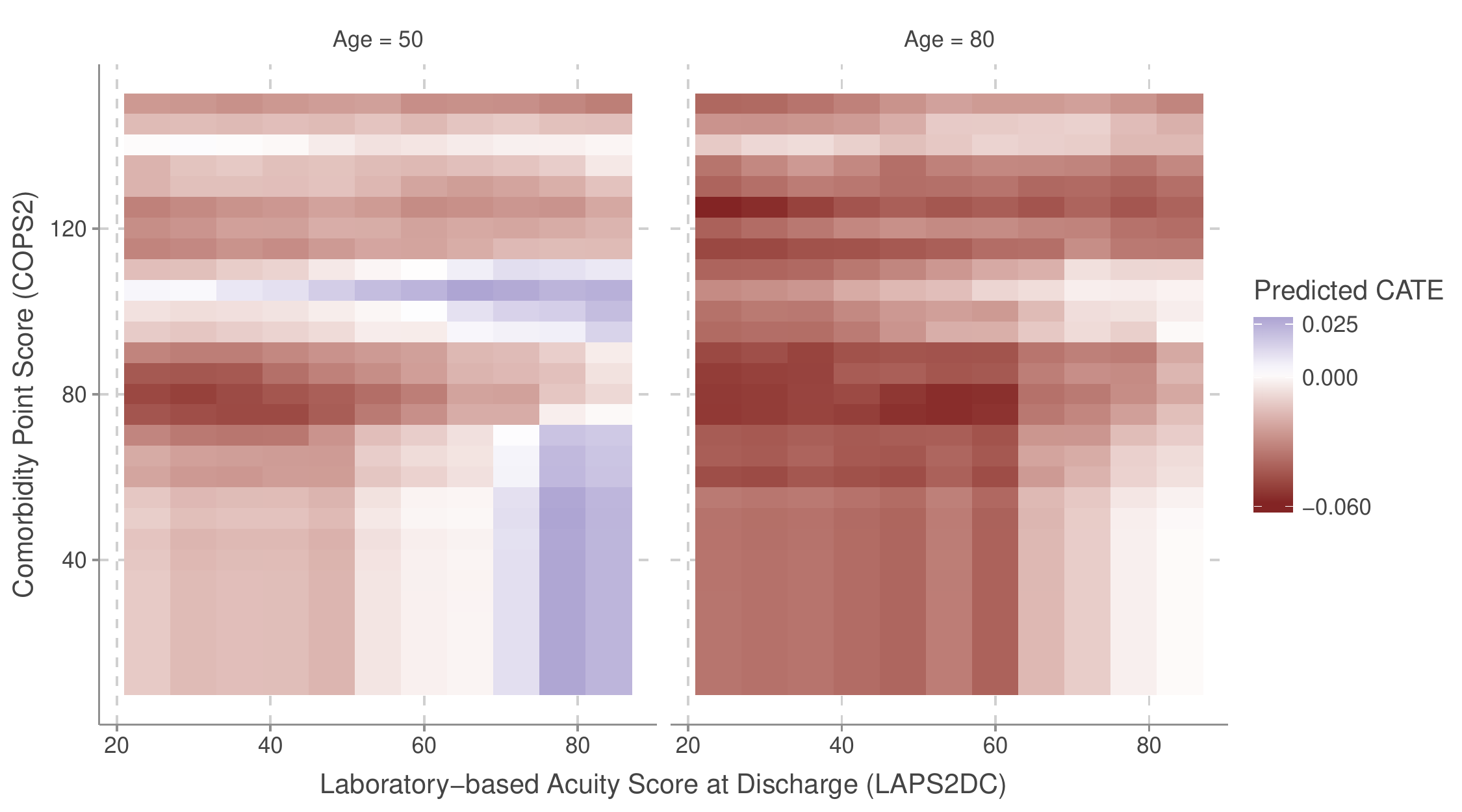}
    \caption{\footnotesize The estimated CATE function as it varies in the dimensions of \texttt{LAPS2DC} and \texttt{COPS2}, for a patient with chronic heart failure at ages 50 and 80. All other continuous variables were fixed at their median, and other categorical variables were fixed at their mode. (The transitions from one cell to another in this figure should not appear smooth; if this is the case, try a different PDF viewer.)}
    \label{fig:cate-chf}
\end{figure}

\subsection{Notional estimates of overall impact under different targeting strategies}

Based on these individual CATE estimates, we compute the potential impacts of several notional targeting strategies using these estimated effects, and not predicted risk, to target the Transitions Program intervention. These impacts are expressed in terms of the number of annual readmissions prevented, and are presented in Table \ref{tab:policies}. These quantities are computed by training a model on all data through December 2017, which is then used to predict effects for patients discharged in 2018. These predicted effects are used to compute the numbers of readmissions prevented and number needed to treat (NNT). We also present the estimated number of interventions required under each strategy.

We first confirm the calibration of the individual effect estimates by taking the same group of patients who were intervened upon under the current risk-based strategy, and use this group to estimate the number of readmissions prevented, with the aim of comparing this number to a previous estimate of the impact of this policy using the average treatment effect \cite{Marafino2020ASystem}. This results in an estimate of 1,246 (95\% confidence interval [CI] 1,110-1,381) readmissions prevented annually, which compares favorably to the previous estimate of 1,210 (95\% CI 990-1,430), representing further evidence that these estimates are well-calibrated. The NNT under both of these these strategies is 33, and the number of individual interventions needed is 39,985. (Table \ref{tab:policies})

Next, computing the impacts of the CATE-based strategies, which target the Transitions Program intervention to the top 10\%, 20\%, and 50\% of each risk ventile, we find that all of these policies are estimated to result in greater potential reductions in the absolute number of readmissions prevented. \ref{tab:policies} The top-10\% strategy may prevent 1,461 (95\% CI 1,294-1,628) readmissions annually, and does so more efficiently, as implied by the NNT of 13. Moreover, the top-20\% strategy requires the same total number of interventions as the existing risk-based strategy (39,648 vs. 39,985), yet is estimated to lead to double the number of annual readmissions prevented, at 2,478 (95\% CI 2,262-2,694). This strategy appears to be much more efficient, as evidenced by its estimated NNT of 16. 

Even under the most expansive strategy, which treats the top 50\% of each risk ventile and requires 250\% the number of total interventions compared to the risk-based strategy, also represents an improvement in the NNT (23 vs. 33). This strategy is estimated to lead to 4,458 (95\% CI 3,925-4,990) readmissions prevented annually, or nearly four times as many as the existing strategy. Finally, we also note that while there appears to exist a tradeoff in terms of absolute impact and efficiency, all CATE-based strategies substantially improved upon the risk-based targeting strategy in terms of the NNT.

\begin{table}[]
\centering
\begin{tabular}{@{}lccc@{}}
\toprule
Treatment strategy & \begin{tabular}[c]{@{}c@{}}Annual readmissions \\ prevented, $n$\end{tabular} & \begin{tabular}[c]{@{}c@{}}Total interventions,\\ $n$\end{tabular} & NNT \\ \midrule
\textbf{Risk-based targeting} & & & \\ 
\quad Target to $ \geq\!\! 25$\% (DiD estimate) & 1,210 (990-1,430) & 39,985 & 33  \\
\quad Target to $ \geq\!\! 25$\% (CF estimate) & 1,246 (1,110-1,381) & 39,985 & 33 \\ \midrule
\textbf{CATE-based targeting} & & & \\ 
\quad Targeting top 10\% & 1,461 (1,294-1,628) & 18,993 & 13 \\
\quad Targeting top 20\% & 2,478 (2,262-2,694) & 39,648 & 16 \\
\quad Targeting top 50\% & 4,458 (3,925-4,990) & 102,534 & 23 \\ \bottomrule
\end{tabular}
  \vspace{0.2em}
 \caption{\footnotesize Estimates of overall impacts of risk-based and notional CATE-based targeting strategies in terms of the annual numbers of readmissions prevented as well as the numbers needed to treat (NNTs) under each targeting strategy, based on the estimates for index admissions in 2018. The first quantity---the difference-in-differences (DiD) estimate---is based on the results of \cite{Marafino2020ASystem}. All quantities are rounded to the nearest integer. Parentheses represent 95\% confidence intervals. Abbreviations: CATE, conditional average treatment effect; DiD, difference-in-differences; CF, causal forest.}
   \label{tab:policies}
\end{table}

\section{Discussion}

In this paper, we have shown the feasibility of estimating individual treatment effects for a comprehensive readmissions prevention intervention using data on over 1.5 million hospitalizations, representing an example of an "impactibility" model \cite{Lewis2010ImpactibilityPrograms}. Even though our analysis used observational data, we found that these individual estimates were well-calibrated, in that none of the individual estimates were greater than the predicted risk. Moreover, these estimates, in aggregate, when used to compute the impact of the risk-based targeting policy, substantially agreed with a separate estimate computed via a difference-in-differences analysis. Notably, our results suggest that strategies targeting similar population health management and quality improvement (QI) interventions based on these individual effects may lead to far greater aggregate benefit compared to targeting based on risk. In our setting, the difference translated to nearly as many as four times the number of readmissions prevented annually over the current risk-based approach.

Our analysis also found both qualitative and quantitative evidence for treatment effect heterogeneity, particularly across levels of predicted risk: the Transitions Program intervention seemed less effective as predicted risk increased. The extent of this mismatch between treatment effect and predicted risk appeared substantial, and may have implications for the design of readmission reduction programs and related population health management programs. Furthermore, our analysis, when stratified by diagnostic supergroup, also appeared to identify distinct subgroups consisting of patients with larger, more negative treatment effects, indicating that the intervention may have been more effective in those subgroups. Patients expected to benefit could be prioritized to receive the intervention, while patients unlikely to benefit could instead receive more targeted care that better meets their needs, including specific subspecialty care, and in some cases, palliative care. More work remains to be done to investigate if these findings hold for other types of preventative interventions, to characterize these subgroups, and to evaluate how to best translate our preliminary findings into practice. Notably, our finding of a risk-treatment effect mismatch is in line with suggestions in the readmissions prevention literature \cite{Finkelstein2020HealthTrial,Steventon2017PreventingRisk,Lindquist2011UnderstandingFactors}. 

To the best of our knowledge, this work is the first to apply causal machine learning together with decision analysis to estimate the treatment effect heterogeneity of a population health management intervention, and as such, represents the first example of an end-to-end "impactibility" model \cite{Lewis2010ImpactibilityPrograms}. Our approach is also notable in that we show how to decouple the causal and predictive aspects of this prediction problem \cite{Kleinberg2015PredictionProblems, Ascarza2018RetentionIneffective}. In particular, our approach was principally inspired by a study of the effectiveness of targeting marketing interventions based on risk by Ascarza \cite{Ascarza2018RetentionIneffective}, as well as previous work on the causal aspects of prediction problems by Kleinberg and co-authors \cite{Kleinberg2015PredictionProblems, Kleinberg2018HumanPredictions}, and others taking causal approaches to prediction problems. \cite{Rubin2006EstimatingMethodology, Blake2015ConsumerExperiment} From a modeling perspective, treating such a problem as purely predictive, as is commonly done in studies developing readmission risk tools \cite{Kansagara2011RiskReview.}, relies on an assumption that may be implausible---specifically that treatment effects correlate with risk. On the other hand, a purely causal approach---one that focuses on modeling treatment effects---fails when confronted with prediction problems that involve resource allocation. Even those patients whose readmissions may be the most "impactible" may not also have the most resource-intensive readmissions. Indeed, this potential oversight exemplifies a form of bias referred to as "omitted payoff bias". \cite{Kleinberg2018HumanPredictions} 

Previous studies have investigated the extent to which preexisting risk assessment tools may already capture a patient's degree of impactibility, by integrating qualitative assessments from nurses and physicians into these models \cite{Flaks-Manov2020PreventingPrediction}. While there seems to be some overlap between predicted risk and impactibility as expressed by these providers, it appears incomplete, and it is not clear how clinician assessments could be integrated into existing models. \cite{Flaks-Manov2020PreventingPrediction} Further afield, performance metrics for predictive models such as the recently proposed "C statistic for benefit" \cite{vanKlaveren2018TheEffects} can assess the ability of such a model to discriminate between patients expected to benefit from a treatment and those who will not. While useful, these metrics do not facilitate assessment of treatment effect heterogeneity. Attempts have been made to model treatment effect heterogeneity among patients undergoing antihypertensive therapy using a similar causal machine learning approach (the X-learner \cite{Kunzel2019MetalearnersLearning}), and also found some extent of mismatch between treatment effects and risk. \cite{Duan2019ClinicalTherapy}

\subsection{The necessity of estimating heterogeneous treatment effects, and not just outcome risk}

Our results highlight the necessity of estimating the treatment effect heterogeneity associated with preventative interventions, particularly at the population scale. A full appraisal of such heterogeneity can aid in targeting these interventions to patients among whom they might be most effective, while targeting intensified versions, or different versions, to patients who may not be expected to benefit from the intervention as originally implemented. As a result, our analyses suggest that up to nearly as four times as many readmissions could be prevented compared to the current state of affairs, while incurring nominal marginal costs in terms of the total number of interventions delivered. Thus, the current risk-centric modeling methodology employed by many hospitals and health systems, as well as by payers, may limit the full potential of these interventions. 

We focus on what could be considered a special case of treatment effect heterogeneity---of that on the absolute risk scale. \cite{Kent2016RiskTrials., Kent2018PersonalizedEffects} Unlike other studies performing similar analyses (e.g. \cite{Kent2008AInfarction} and others cited in \cite{Kent2016RiskTrials.}) finding that a small group of high-risk patients accounted for most of the aggregate benefits, we instead found that the effect of the Transitions Program intervention were largest in the most numerous subgroup of patients, namely those at relatively low to moderate risk. For high-risk patients, the intervention appeared to be less effective, which is in line with the growing body of evidence showing similar findings \cite{Finkelstein2020HealthTrial,Steventon2017PreventingRisk,Lindquist2011UnderstandingFactors,Rich1993PreventionStudy.}. An open question is the extent to which these high-risk/ineffective patients are qualitatively distinct, to the point that they should be considered "futile", as some have proposed \cite{Lewis2010ImpactibilityPrograms}, or if they may be amenable to a more intensified version of the intervention or an entirely different approach altogether.

A notable finding is that some patients had a predicted CATE greater than zero, indicating that the Transitions Program intervention may have encouraged them to return to the hospital. In other settings, a positive treatment effect would often be interpreted as harm, and suggest that the treatment be withheld from these patients. However, we argue that our finding does not readily admit such an interpretation. To see why, we note that this subgroup of patients with a positive CATE appeared to be those who were more acutely ill at discharge, as evidenced by their higher \texttt{LAPS2DC} scores (Figure \ref{fig:cate-chf}). In light of this finding, an alternative interpretation of these weakly positive effects is that they represent readmissions which may have been necessary, and which perhaps may have been facilitated by aspects of the Transitions Program intervention, including instructions to patients outlining the circumstances (e.g. new or worsening symptoms) under which they should return to the hospital. This finding holds particular relevance given increasing concern that readmission prevention programs, in responding to the incentives of the HRRP, may be reducing 30-day hospitalization rates at the expense of increased short- and long-run mortality. \cite{Wadhera2018AssociationPneumonia, Fonarow2017TheReconsider, Gupta2018ThePolicy} 

Moreover, this finding also suggests that the CATE estimates may be insufficient to capture the full impact of the Transitions Program intervention on patient outcomes, meaning that the estimated effect of the intervention on readmission alone may not represent a sufficient basis for future targeting strategies. It is plausible that intervening in a patient with a positive estimated effect may be warranted if the readmission would have a positive effect on other outcomes, despite the current emphasis of value-based purchasing programs on penalizing excess 30-day readmissions. For example, in fiscal year 2016, the maximum penalty for excess 30-day mortality was 0.2\% of a hospital's diagnosis-related group (DRG) payments under the Hospital Value-Based Purchasing program, while the maximum penalty for excess 30-day readmission was 3.0\% of DRG payments under the HRRP. \cite{Abdul-Aziz2017AssociationMortality} A more holistic targeting strategy would incorporate estimates of the intervention's effect on short- and long-run mortality and other outcomes, and explicitly include these quantities when computing expected utility. Selecting patients for treatment can then be formulated as an optimization problem that attempts to balance regulatory incentives, organizational priorities, patient welfare, and resource constraints.

\subsection{Causal aspects of prediction problems}

Our approach, in contrast to many studies which have developed readmission risk prediction models \cite{Kansagara2011RiskReview.,Bayati2014Data-drivenStudy.,Bates2014BigPatients}, instead focuses on the causal aspects of the readmissions prediction problem. Our analyses emphasize modeling treatment effect heterogeneity of the Transitions Program intervention and on closing the loop from prediction to decision. However, more work remains to formalize criteria that more clearly delineate the roles of causal inference and prediction in these prediction problems, and the extent to which either is necessary or sufficient for a given  problem. 

As explored by \cite{Bayati2014Data-drivenStudy.} and \cite{Kleinberg2018HumanPredictions}, fully operationalizing a prediction model entails the following steps:

\begin{equation*}
    \text{data} \Rightarrow \text{prediction} \Rightarrow \text{decision}.
\end{equation*}

Historically, many machine learning studies, including those in medicine, have emphasized the $\textit{data} \Rightarrow \textit{prediction}$ link, while neglecting the $\textit{prediction} \Rightarrow \textit{decision}$ link. These studies often evaluate model quality on the basis of performance metrics, including the area under the reciever operating characteristic curve (AUROC or AUC) or C-statistic, the area under the precision-recall curve (AUPRC), and other measures of accuracy. A final model is chosen from among candidate models because it maximizes the value of the metric of interest, with the implicit hope that it also maximizes its utility in a real-world setting were it to be so applied. However, this approach conflates prediction with classification---the risks of doing so which are clear, though perhaps underappreciated \cite{Harrell2019ClassificationThinking}---and hence also conflates prediction quality with decision quality. 

Indeed, it is conceivable (though not likely) that even an intervention based on a perfectly accurate prediction model---for example, one that correctly identified all readmissions---could result in no net benefit, if the intervention proved ineffective for all the "high-risk" patients whose readmissions were identified. This contrasts with the setting of many classification problems (which we emphasize again are distinct from prediction problems), including, for example, image classification. In these cases, the utility function of classification is expressible in terms of the performance metric of interest alone, such as the accuracy or $F_1$ score. For a given choice of metric, the utility of an image classifier is strictly increasing in that metric, whereas in our hypothetical readmission prediction model, even perfect "classification" is not sufficient. This mismatch between predictive performance and utility is less pronounced with a less severe extent of treatment effect heterogeneity, but our hypothetical example highlights the importance of making the distinction.

To be sure, not all prediction problems decompose neatly into causal and predictive components. Some such problems are purely causal, while others can be solved with prediction alone. Models developed for risk adjustment, for example, represent an example of the latter types of problems, as in those contexts there is no link between an individual prediction and a decision. Rather, the predictions of risk adjustment are used in aggregate; for example, to obtain observed-to-expected ratios of outcomes for quality measures, or to adjust capitation payments according to case mix. \cite{MedPAC2016MedicareSystem} Another class of prediction models that can be viewed as solving purely predictive problems are those used for clinical decision support \cite{Chen2017MachineExpectations}, including the APACHE ICU mortality model \cite{Knaus1985APACHESystem}. 

These decision support systems can be used in the context of an individual patient at the point of care to inform decision-making. For example, the APACHE model can be used to track the response of a patient in the intensive care unit (ICU) to treatment, or an objective measure to help decide when palliative care may be necessary. \cite{Knaus2002APACHEReflections} But it is not used in isolation to decide when to start and stop an intervention, unlike the prediction model used to enroll patients in the Transitions Program which we studied in this paper. Instead, the physician is meant to integrate APACHE model output with other information to arrive at a decision, which may incorporate other clinical data not captured in the model, resource constraints, or even family preferences. This is an significant distinction: the view of the APACHE model as solving a purely predictive problem implicitly off-loads the burden of decision-making, and thus of conceptualizing an utility function, to the physician. Without a human (or physician) in the loop, the utility function must be made more explicit in order to maximize aggregate utility, as in the framework we describe in this paper.

On the other hand, a setting where the link between prediction and decision is more explicit is that of resource allocation problems. A well-known example involves the MELD score \cite{Kamath2001ADisease}, which predicts 3-month mortality in liver failure and is used to prioritize patients for liver transplant. The prediction problem can again be classed as purely predictive, because patients with higher MELD scores (and hence higher 3-month mortality) benefit most: allocating donor livers to the patients with the highest MELD scores yields more years of life saved, in aggregate. A similar insight was employed by Kleinberg and co-authors \cite{Kleinberg2015PredictionProblems}, who investigated a different resource allocation problem---that of allocating joint replacement surgeries. A patient who receives a joint replacement will not see the full benefit until at least a year post-surgery, due to the time required for recovery and physical therapy in the interim. 

Hence, patients who receive a joint replacement, but die within a year post-surgery, do not benefit from surgery. With this in mind, the problem of allocating joint replacements can be cast as a purely predictive one by developing a prediction model for mortality at 1 year post-surgery. Surgeries can then be allocated to patients in need below a certain risk threshold. Unlike with the MELD score, the authors found that risk did not appear to correlate with benefit as measured by claims for physical therapy, joint injections, and physician visits pre-surgery. If the riskiest patients would otherwise have derived the most benefit from surgery, this would constitute an example of "omitted payoff bias", requiring a different utility function that took the payoff of decreased medical need (presumably corresponding to less severe symptoms) into account in order to assign patients to surgery. \cite{Kleinberg2015PredictionProblems}

We close this subsection with an example of a prediction problem that appears purely predictive, but is in fact largely, and perhaps purely, causal. Consider the problem of identifying hospitalized patients who would benefit from a palliative care consultation, which has attracted considerable attention over the last two decades as the need for palliative care has grown. \cite{Weissman2011IdentifyingCare.} Holistic criteria have been proposed to identify these patients, spanning domains including mortality risk in the next 12 months, the presence of refractory symptoms, the patient's level of social support, and inability to carry out activities of daily living---which constitute a selection of the 13 indicators outlined in \cite{Weissman2011IdentifyingCare.}. 

Recently, attempts have also been made to reduce the problem of identifying patients who might benefit from palliative care to one of modeling 12-month mortality from hospital admission, e.g., \cite{Avati2018ImprovingLearning}. Notwithstanding the causal issues involved in defining this outcome of interest retrospectively and applying it prospectively (e.g., see \cite{Einav2018PredictiveLife} for a related discussion), this approach is also problematic in that it is unlikely to reliably identify patients who will actually benefit from palliative care, because it all omits other indicators of palliative care need beyond 12-month mortality. A palliative care consultation is a complex intervention which will likely have highly heterogeneous treatment effects on quality of life---which is the real outcome of interest---and it is unclear if these effects do in fact correlate with 12-month mortality risk.

In fact, solving this proxy problem will likely identify many patients who will die in the short run, but who may not be amenable to palliative care. This includes patients in the ICU, for whom providing the full spectrum of palliative care can be challenging, due to sedation, the use of mechanical ventilation and other invasive interventions, environmental light and noise, and limits on family visitation. \cite{Cook2014DyingUnit} Conversely,  this approach would also miss many patients who are at low mortality risk, but who may otherwise benefit from palliative care early in their disease course. Indeed, guidelines recommend initiating palliative care early in patients with cancer, for example---with some trials finding that initiation at as early as the time of diagnosis may be most effective. \cite{Howie2013EarlyImplications}

\subsection{New processes for predictive algorithm-driven intervention deployments within learning health systems}

Our findings could be used to retarget the Transitions Program intervention prospectively to patients who are most expected to benefit, rather than those at highest risk, resulting in larger aggregate benefit in terms of the number of readmissions prevented. Similarly, our approach could be used to retarget other population health management and QI interventions which are often deployed wholesale or using risk assessment tools. However, as we mention, these individual estimates were derived from observational data, and not from data generated via a randomized experiment---the latter type of data which represent the ideal substrate for estimating treatment effects insofar as randomization is able to mitigate the effects of confounding. \cite{Kent2018PersonalizedEffects} Furthermore, our approach requires interventional data, unlike those used to develop more traditional risk assessment tools, which are often based on retrospective data. Hence, to implement our approach, as an alternative to risk tool-driven approaches, may require rethinking how these predictive algorithm-driven interventions (or "prediction-action dyads", cf. \cite{Liu2019TheHealthcare}) are deployed within health systems, particularly in relation to existing digital infrastructure and institutional oversight processes. We outline several starting points for doing so below. 

One option is to first deploy a new predictive algorithm-driven intervention as part of a simple two-arm randomized trial which compares that intervention to usual care. This represents a pilot phase, generating data that are used to derive an impactibility model, along with a payoff model if necessary for the prediction problem. Following this pilot phase, two paths are possible: 1) based on this impactibility model, the intervention could be re-targeted to the patients most expected to benefit; or 2) alternatively, carrying out another two-arm randomized trial comparing risk-based to impactibility-based targeting. In the latter choice, patients would be randomized to either of the risk or impactibility arms, and based on their covariates would either receive or not receive the intervention according to their risk or benefit estimate. Based on the results of this second trial, whichever targeting approach proved more effective could then be put into use.

Another option, if a predictive algorithm-driven intervention based on risk scores is already in use, does away with the pilot randomized trial. Instead, the first step is to derive a impactibility model from observational data, as we did in this study. Then, if the goal is to retarget this intervention to include more patients at lower risk---for example, patients between 10 and 25\% predicted risk---risk-based and impactibility-based targeting could be compared to each other in a three-arm randomized trial. The first arm consists of risk-based targeting using the current risk threshold; the second arm, impactibility-based targeting; and the third arm, risk-based targeting to patients in the 10-25\% risk subgroup. Finally, cluster-randomized trial designs based on the regression discontinuity design would allow treatment effect heterogeneity across levels of risk to be characterized, enabling the intervention to be retargeted to the groups among which it is most effective; we hope to investigate these designs in future work.

These proposed options represent major shifts from how deployments of predictive algorithm-driven interventions are usually carried out in health systems. New institutional overnight processes \cite{Faden2013AnEthics}, digital infrastructure, and statistical methodology would be required in order to realize the full potential of these approaches. Many such interventions, if deemed to create only minimal risk, fall under the umbrella of routine quality improvement (QI) studies, which exempts them from ongoing independent oversight \cite{Finkelstein2015OversightResearch}. However, incorporating randomization, as we do in the options we describe above, shifts these interventions from the category of routine QI to non-routine QI or research. These two categories of studies often require independent oversight by, for example, an institutional review board (IRB), and may need to incorporate additional ethical considerations, e.g., requiring informed consent. It is possible that a categorization of a deployment process like the ones we describe above as non-routine QI or research could impede their being carried out, as has occurred in previous QI work. \cite{Baily2006SpecialSafety} However, similar attempts at deploying QI interventions as a part of rapid-cycle randomized experiments have escaped such categorization \cite{Horwitz2019CreatingTesting}, although standards will likely vary from institution to institution.

Moreover, these new deployment processes we envision will also require new digital infrastructure to implement these models and provision their linked interventions as a part of randomized trials and to monitor their impacts in real-time (while accounting for early stopping and other biases), akin to how A/B testing is usually carried out by many Web-facing companies today \cite{Kohavi2013OnlineScale}. Another sticking point is the lack of established analogues of power analyses for causal forests and other methods. Without these analogues, it is not clear how to set the necessary sample size for a pilot randomized trial like the one we describe above, although it could possibly be determined via simulation. The investments required to establish these platforms may be considerable, and it remains unclear whether the costs in fact do outweigh the rewards, particularly when one also accounts for the opportunity costs associated with experimentation. However, the potential impact, both on patient outcomes and in terms of the knowledge generated, could be substantial, and merit further investigation.

\subsection{Limitations}

This study has several limitations. First, as this study is observational in nature, our analysis necessarily relies on certain assumptions, which, while plausible, are unverifiable. The unconfoundedness assumption that we make presumes no unmeasured confounding, and cannot be verified through inspection of the data nor via statistical tests. It is possible that our results, particularly the notional estimates of overall impact in terms of the numbers of readmissions prevented annually under the CATE-based targeting strategies, are biased. However, it appears that the magnitude of unobserved bias would have to be large to negate our results, particularly our estimates of the potential impacts of the CATE-based strategies. Furthermore, because the causal forest performs local linear estimation, the assumptions that we make with respect to the risk threshold can be considered modular in the sense that if they fail for the subgroup of patients with a predicted risk of below 25\% across both the pre- and post-implementation periods (which we anticipate would be most likely), they would still hold for the $\geq \!\!25$\% risk subgroup.

Moreover, we again note that the CATEs appeared well-calibrated in at least two aspects: the number of readmissions prevented under the risk-based targeting strategy, as estimated using the predicted CATEs, agreed with that estimated via a separate analysis; and the predicted CATEs were never larger than the predicted risk for any individual. Furthermore, this limitation becomes moot if the data are generated by a randomized experiment. We outline several approaches for how existing processes of deploying predictive algorithm-driven interventions could be redesigned to incorporate randomization in order to iteratively refine these interventions through improved targeting. 

Second, these estimates of benefit must be computed with respect to an intervention, which cannot be made with historical data alone, as is often done when developing risk assessment tools. As such, the necessary data must be generated, whether through wholesale deployments of an intervention (as in many QI studies) or via randomized experiments. The latter provide a better basis for these analyses, but are costly and require additional infrastructure, and may be subject to more institutional oversight. Third, our simplifying assumption with respect to the nature of the payoffs may not have resulted in a targeting strategy that was in fact optimal, but we hope to explore in future work the feasibility of predicting payoffs using supervised machine learning in parallel with treatment effects within this framework, and observing how the resulting targeting strategies change. 

Third, although we incorporated it into our analysis, the \texttt{HCUPSGDC} variable is not uniformly always available at discharge. From the point of a view of a retrospective analysis that principally seeks to characterize treatment effect heterogeneity, this does not constitute a limitation. However, this consideration may preclude the application of this impactibility model prospectively in its current form, but it is possible that the patient's problem list at discharge could be used to infer the value of this variable, as there are only 25 categories used in our analysis. Finally, our results may not be applicable to all settings, particularly hospitals and health systems without a high degree of integration. However, the framework we outline here is agnostic as to application as well as the type of causal machine learning model used, and could be applied to resource allocation and other prediction problems more generally.

\section{Conclusion}

Causal machine learning can be used to identify preventable hospital readmissions, if the requisite interventional data are available. Moreover, our results point to a mismatch between readmission risk and treatment effect, which is consistent with suggestions in prior work. In our setting, the extent of this mismatch was considerable, suggesting that many preventable readmissions may be being "left on the table" with current risk modeling methodology. Our proposed framework is also generalizable to the study of a variety of population health management and quality improvement interventions driven by predictive models, as well as of algorithm-driven interventions in a range of settings outside of healthcare.

\newpage

\section*{Acknowledgements}

The authors are immensely grateful to Colleen Plimier of the Division of Research, Kaiser Permanente Northern California, for assistance with data preparation, as well as to Dr. Tracy Lieu, also of the Division of Research, for reviewing the manuscript. In addition, the authors wish to thank  Minh Nguyen, Stephen Pfohl, Scotty Fleming, and Rachael Aikens for helpful feedback on earlier versions of this work. 

Mr. Marafino was supported by a predoctoral fellowship from the National Library of Medicine of the National Institutes of Health under Award Number T15LM007033, as well as by funding from a Stanford School of Medicine Dean's Fellowship. Dr. Baiocchi was also supported by grant KHS022192A from the Agency for Healthcare Research and Quality. Dr. Vincent Liu was also supported by NIH grant R35GM128672 from the National Institute of General Medical Sciences. Portions of this work were also funded by The Permanente Medical Group, Inc., and Kaiser Foundation Hospitals, Inc. The content is solely the responsibility of the authors and does not necessarily represent the official views of the National Institutes of Health. The authors have no conflicts of interest to disclose. The funders played no role in the study design, data collection, analysis, reporting of the data, writing of the report, nor the decision to submit the article for publication. 

\section*{Appendix A}
\setcounter{table}{0}
\renewcommand{\thetable}{A\arabic{table}}

\begin{table}[H]
\centering
\resizebox{\textwidth}{!}{%
\begin{tabular}{@{}lccccc@{}}
\toprule
 & Total & Pre-implementation & Post-implementation & \multicolumn{1}{l}{$p$-value} & \multicolumn{1}{l}{SMD} \\ \midrule
Hospitalizations, $n$ & 1,584,902 & 1,161,452 & 423,450 & --- & ---  \\
Patients, $n$ & 753,587 & 594,053 & 266,478 & --- & ---  \\
Inpatient (\%) & 82.8 (69.7-90.6) & 84.4 (73.1-90.7) & 78.5 (57.7-90.3) & $<0.0001$ & -0.151 \\
Observation (\%) & 17.2 (9.4-30.3) & 15.6 (9.3-26.9) & 21.5 (9.7-42.3) & $<0.0001$ & 0.151 \\
Inpatient stay $<24$ hours & 5.2 (3.3-6.7) & 5.1 (3.8-6.5) & 5.6 (1.8-9.1) & $<0.0001$ & 0.041 \\
Transport-in & 4.5 (1.4-8.7) & 4.5 (1.7-8.8) & 4.5 (0.4-8.4) & 0.56 & -0.001 \\
Age, mean (years) & 65.3 (62.2-69.8) & 65.1 (61.9-69.6) & 65.8 (62.8-70.4) & $<0.0001$ & 0.038 \\
Male gender (\%) & 47.5 (43.4-53.8) & 47.0 (42.4-53.5) & 48.9 (45.3-54.9 & $<0.0001$ & 0.037 \\
KFHP membership (\%) & 93.5 (75.3-97.9) & 93.9 (80.0-98.0) & 92.5 (61.7-97.6) & $<0.0001$ & -0.052 \\
Met strict membership definition (\%) & 80.0 (63.4-84.9) & 80.6 (67.6-85.5) & 78.5 (51.3-83.8) & $<0.0001$ & -0.053 \\
Met regulatory definition (\%) & 61.9 (47.2-69.7) & 63.9 (50.2-72.2) & 56.5 (38.7-66.6) & $<0.0001$ & -0.152 \\
Admission via ED (\%) & 70.4 (56.7-82.0) & 68.9 (56.0-80.3) & 74.4 (58.4-86.6) & $<0.0001$ & 0.121 \\
Charlson score, median (points) & 2.0 (2.0-3.0) & 2.0 (2.0-3.0) & 2.0 (2.0-3.0) & $<0.0001$ & 0.208 \\
Charlson score $\geq 4$ (\%) & 35.2 (29.2-40.7) & 33.2 (28.2-39.8) & 40.9 (33.0-46.2) & $<0.0001$ & 0.161 \\
COPS2, mean (points) & 45.6 (39.1-52.4) & 43.5 (38.4-51.5) & 51.2 (39.7-55.8) & $<0.0001$ & 0.159 \\
COPS2 $\geq 65$ (\%) & 26.9 (21.5-32.0) & 25.3 (21.0-31.6) & 31.1 (22.5-35.4) & $<0.0001$ & 0.129 \\
Admission LAPS2, mean (points) & 58.6 (48.0-67.6) & 57.6 (47.4-65.8) & 61.3 (50.2-72.8) & $<0.0001$ & 0.092 \\
Discharge LAPS2, mean (points) & 46.7 (42.5-50.8) & 46.3 (42.5-50.8) & 47.6 (42.3-52.9) & $<0.0001$ & 0.039 \\
LAPS2 $\geq 110$ (\%) & 12.0 (7.8-16.0) & 11.6 (7.5-15.2) & 12.9 (8.3-18.4) & $<0.0001$ & 0.039 \\
Full code at discharge (\%) & 84.4 (77.3-90.5) & 84.5 (77.7-90.5) & 83.9 (75.9-90.5) & $<0.0001$ & -0.016 \\
Length of stay, days (mean) & 4.8 (3.9-5.4) & 4.9 (3.9-5.4) & 4.7 (3.9-5.6) & $<0.0001$ & -0.034 \\
Discharge disposition (\%) &  &  &  &  & 0.082 \\
\quad To home & 72.7 (61.0-86.2) & 73.3 (63.9-85.9) & 71.0 (52.1-86.9) & $<0.0001$ &  \\
\quad Home Health & 16.1 (6.9-23.3) & 15.2 (6.9-22.6) & 18.5 (7.0-34.5) & $<0.0001$ &  \\
\quad Regular SNF & 9.9 (5.9-14.3) & 10.0 (6.0-15.2) & 9.5 (5.6-12.4) & $<0.0001$ &  \\
\quad Custodial SNF & 1.3 (0.7-2.5) & 1.5 (0.8-2.7) & 0.9 (0.4-1.8) & $<0.0001$ &  \\
Hospice referral (\%) & 2.6 (1.7-4.4) & 2.6 (1.7-4.6) & 2.7 (1.5-4.0) & $<0.0001$ & 0.007 \\ \midrule
\textbf{Outcomes}&  &  &  &  &  \\
\quad Inpatient mortality (\%) & 2.8 (2.1-3.3) & 2.8 (2.1-3.3) & 2.8 (1.8-3.3) & 0.17 & -0.003 \\
\quad 30-day mortality (\%) & 6.0 (4.0-7.3) & 6.1 (4.1-7.6) & 5.9 (3.9-6.8) & $<0.0001$ & -0.006 \\
\quad Any readmission (\%) & 14.5 (12.7-17.2) & 14.3 (12.3-17.3) & 15.1 (13.3-17.0) & $<0.0001$ & 0.021 \\
\quad Any non-elective readmission (\%) & 12.4 (10.4-15.4) & 12.2 (10.2-15.5) & 13.1 (10.8-15.4) & $<0.0001$ & 0.029 \\
\quad Non-elective inpatient readmission (\%) & 10.5 (8.2-12.6) & 10.4 (8.1-12.8) & 10.8 (8.6-12.9) & $<0.0001$ & 0.012 \\
\quad Non-elective observation readmission (\%) & 2.4 (1.4-3.7) & 2.2 (1.2-3.4) & 3.0 (1.9-5.6) & $<0.0001$ & 0.049 \\
\quad 30-day post-discharge mortality (\%) & 4.0 (2.6-5.2) & 4.1 (2.7-5.4) & 3.9 (2.3-4.9) & $<0.0001$ & -0.007 \\ 
\quad Composite outcome (\%) & 15.2 (12.9-18.8) & 15.0 (12.9-19.1) & 15.8 (13.3-18.0) & $<0.0001$ & 0.023 \\ \bottomrule
\end{tabular}}
 \caption{Characteristics of the cohort, including both index and non-index stays. Notably, comparing pre- to post-implementation, hospitalized patients were older, and tended to have higher comorbidity burden (higher COPS2) as well as a higher acuity of illness at admission (higher LAPS2). The use of observation stays also increased. These differences reflect a broader trend towards the pool of potential inpatient admissions becoming more and more ill over the decade from 2010, in large part due to the effectiveness of outpatient preventative care processes at KPNC, as well as of programs providing care outside of the hospital setting as an alternative to admission. Otherwise, care patterns did not substantially change, as evidenced by, e.g., transports-in, Kaiser Foundation Health Plan (KFHP) membership status, and discharge disposition mix, all of which had standardized mean differences (SMDs) $<0.1$. In large cohorts such as this one, SMDs can be a better guide to detecting covariate imbalances or differences between groups, owing to the effects of large sample sizes. Finally, as a consequence of increased comorbidity burden and admission acuity, and despite the implementation of the Transitions Program, rates of readmission and of the composite outcome increased from pre- to post-implementation. Abbreviations: SMD, standardized mean difference; KFHP, Kaiser Foundation Health Plan; LAPS2,  Laboratory-based Acute Physiology Score, version 2; COPS2, COmorbidity Point Score, version 2; SNF, skilled nursing facility.}
   \label{tab:a1}

\end{table}

\section*{Appendix B}
\setcounter{table}{0}
\renewcommand{\thetable}{B\arabic{table}}

\begin{table}[H]
\centering
\resizebox{\textwidth}{!}{%
\begin{tabular}{@{}lc@{}}
\toprule
Supergroup name (\texttt{HCUPSGDC}) & \begin{tabular}[c]{@{}c@{}}Clinical Classification Software (CCS) \\ category code(s)\end{tabular} \\ \midrule
Acute CVD & 109 \\
AMI & 100 \\
CAP & 122 \\
Cardiac arrest & 107 \\
CHF & 108 \\
Coma; stupor; and brain damage & 85 \\
Endocrine \& related conditions & 48-51, 53, 54, 56, 58, 200, 202, 210, 211 \\
Fluid and electrolyte disorders & 55 \\
GI bleed & 153 \\
Hematologic conditions & 59-64 \\
Highly malignant cancer & 17, 19, 27, 33, 35, 38-43 \\
Hip fracture & 226 \\
Ill-defined signs and symptoms & 250-253 \\
Less severe cancer & 11-16, 18, 20-26, 28-32, 34, 36, 37, 44-47, 207 \\
Liver and pancreatic disorders & 151, 152 \\
Miscellaneous GI conditions & 137-140, 155, 214 \\
Miscellaneous neurological conditions & 79-84, 93-95, 110-113, 216, 245, 653 \\
Miscellaneous surgical conditions & 86-89, 91, 118-121, 136, 142, 143, 167, 203, 204, 206, 208, 209, 212, 237, 238, 254, 257 \\
Other cardiac conditions & 96-99, 103-105, 114, 116, 117, 213, 217 \\
Other infectious conditions & 1, 3-9, 76-78, 90, 92, 123-126, 134, 135, 148, 197-199, 201, 246-248 \\
Renal failure (all) & 156, 157, 158 \\
Residual codes & 259 \\
Sepsis & 2 \\
Trauma & 205, 225, 227-236, 239, 240, 244 \\
UTI & 159 \\ \bottomrule
\end{tabular}}
\caption{List of Clinical Classification Software (CCS)-defined supergroups and their CCS codes used in this study. These supegroups represent levels of the covariate \texttt{HCUPSGDC}. More details on the CCS codes themselves, as well as mappings to their component ICD codes, can be found at \url{www.ahrq.gov/data/hcup}.}
\label{tab:b1}
\end{table}

\newpage

\printbibliography

@article{Athey,
    title = {{Estimating Treatment Effects with Causal Forests: An Application}},
    year = {2019},
    journal = {arXiv},
    author = {Athey, Susan and Wager, Stefan},
    pages = { https://arxiv.org/pdf/1902.07409.pdf},
    url = {https://arxiv.org/pdf/1902.07409.pdf},
    arxivId = {1902.07409v1}
}

@article{AtheyGRF,
    title = {{Generalized Random Forests}},
    year = {2018},
    author = {Athey, Susan and Tibshirani, Julie and Wager, Stefan},
    arxivId = {arXiv:1610.01271v4}
}

@article{Chernozhukov2017,
    title = {{Generic Machine Learning Inference on Heterogenous Treatment Effects in Randomized Experiments}},
    year = {2017},
    author = {Chernozhukov, Victor and Demirer, Mert and Duflo, Esther and Fernandez-Val, Ivan},
    month = {12},
    url = {http://arxiv.org/abs/1712.04802},
    arxivId = {1712.04802}
}

@article{Lewis2010ImpactibilityPrograms,
    title = {{"Impactibility models": Identifying the subgroup of high-risk patients most amenable to hospital-avoidance programs}},
    year = {2010},
    journal = {Milbank Quarterly},
    author = {Lewis, Geraint H.},
    number = {2},
    month = {6},
    pages = {240--255},
    volume = {88},
    doi = {10.1111/j.1468-0009.2010.00597.x},
    issn = {0887378X}
}

@article{Marafino2020ASystem,
    title = {{A Comprehensive Readmissions Prevention Intervention Enabled by Predictive Analytics in an Integrated Health System}},
    year = {2020},
    journal = {to appear},
    author = {Marafino, Ben J and Escobar, Gabriel J and Liu, Vincent X and Baiocchi, Mike and Schuler, Alejandro}
}

@article{Kamath2001ADisease,
    title = {{A model to predict survival in patients with end-stage liver disease}},
    year = {2001},
    journal = {Hepatology},
    author = {Kamath, Patrick S. and Wiesner, Russell H. and Malinchoc, Michael and Kremers, Walter and Therneau, Terry M. and Kosberg, Catherine L. and D’amico, Gennaro and Dickson, E. Rolland and Kim, W. Ray},
    number = {2},
    pages = {464--470},
    volume = {33},
    doi = {10.1053/jhep.2001.22172},
    issn = {02709139},
    pmid = {11172350}
}

@article{Kent2008AInfarction,
    title = {{A Percutaneous Coronary Intervention-Thrombolytic Predictive Instrument to Assist Choosing Between Immediate Thrombolytic Therapy Versus Delayed Primary Percutaneous Coronary Intervention for Acute Myocardial Infarction}},
    year = {2008},
    journal = {American Journal of Cardiology},
    author = {Kent, David M. and Ruthazer, Robin and Griffith, John L. and Beshansky, Joni R. and Concannon, Thomas W. and Aversano, Thomas and Grines, Cindy L. and Zalenski, Robert J. and Selker, Harry P.},
    number = {6},
    month = {3},
    pages = {790--795},
    volume = {101},
    url = {http://www.ncbi.nlm.nih.gov/pubmed/18328842},
    doi = {10.1016/j.amjcard.2007.10.050},
    issn = {00029149}
}

@techreport{AgencyforHealthcareResearchandQuality2001AHRQConditions.,
    title = {{AHRQ quality indicators—guide to prevention quality indicators: hospital admission for ambulatory care sensitive conditions.}},
    year = {2001},
    author = {{Agency for Healthcare Research and Quality}},
    pages = {Publication No: 02-R0203}
}

@article{Faden2013AnEthics,
    title = {{An Ethics Framework for a Learning Health Care System: A Departure from Traditional Research Ethics and Clinical Ethics}},
    year = {2013},
    journal = {Hastings Center Report},
    author = {Faden, Ruth R. and Kass, Nancy E. and Goodman, Steven N. and Pronovost, Peter and Tunis, Sean and Beauchamp, Tom L.},
    number = {s1},
    month = {1},
    pages = {S16-S27},
    volume = {43},
    publisher = {John Wiley {\&} Sons, Ltd},
    url = {http://doi.wiley.com/10.1002/hast.134},
    doi = {10.1002/hast.134},
    issn = {00930334}
}

@misc{Knaus2002APACHEReflections,
    title = {{APACHE 1978-2001: The development of a quality assurance system based on prognosis: Milestones and personal reflections}},
    year = {2002},
    booktitle = {Archives of Surgery},
    author = {Knaus, W. A.},
    number = {1},
    month = {1},
    pages = {37--41},
    volume = {137},
    publisher = {American Medical Association},
    doi = {10.1001/archsurg.137.1.37},
    issn = {00040010}
}

@article{Knaus1985APACHESystem,
    title = {{APACHE II: a severity of disease classification system}},
    year = {1985},
    journal = {Critical Care Medicine},
    author = {Knaus, W A and Draper, E A and Wagner, D P and Zimmerman, J E},
    number = {10},
    pages = {818--829},
    volume = {13},
    url = {http://www.ncbi.nlm.nih.gov/pubmed/3928249},
    issn = {0090-3493},
    language = {eng}
}

@article{Abdul-Aziz2017AssociationMortality,
    title = {{Association between Medicare hospital readmission penalties and 30-day combined excess readmission and mortality}},
    year = {2017},
    journal = {JAMA Cardiology},
    author = {Abdul-Aziz, Ahmad A. and Hayward, Rodney A. and Aaronson, Keith D. and Hummel, Scott L.},
    number = {2},
    month = {2},
    pages = {200--203},
    volume = {2},
    publisher = {American Medical Association},
    doi = {10.1001/jamacardio.2016.3704},
    issn = {23806591}
}

@article{Berkowitz2018AssociationJ-CHiP,
    title = {{Association of a Care Coordination Model With Health Care Costs and Utilization: The Johns Hopkins Community Health Partnership (J-CHiP)}},
    year = {2018},
    journal = {JAMA network open},
    author = {Berkowitz, Scott A. and Parashuram, Shriram and Rowan, Kathy and Andon, Lindsay and Bass, Eric B. and Bellantoni, Michele and Brotman, Daniel J. and Deutschendorf, Amy and Dunbar, Linda and Durso, Samuel C. and Everett, Anita and Giuriceo, Katherine D. and Hebert, Lindsay and Hickman, Debra and Hough, Douglas E. and Howell, Eric E. and Huang, Xuan and Lepley, Diane and Leung, Curtis and Lu, Yanyan and Lyketsos, Constantine G. and Murphy, Shannon M.E. and Novak, Tracy and Purnell, Leon and Sylvester, Carol and Wu, Albert W. and Zollinger, Ray and Koenig, Kevin and Ahn, Roy and Rothman, Paul B. and Brown, Patricia M.C.},
    number = {7},
    month = {11},
    pages = {e184273},
    volume = {1},
    publisher = {NLM (Medline)},
    doi = {10.1001/jamanetworkopen.2018.4273},
    issn = {25743805}
}

@article{Wadhera2018AssociationPneumonia,
    title = {{Association of the Hospital Readmissions Reduction Program With Mortality Among Medicare Beneficiaries Hospitalized for Heart Failure, Acute Myocardial Infarction, and Pneumonia}},
    year = {2018},
    journal = {JAMA},
    author = {Wadhera, Rishi K. and Joynt Maddox, Karen E. and Wasfy, Jason H. and Haneuse, Sebastien and Shen, Changyu and Yeh, Robert W.},
    number = {24},
    month = {12},
    pages = {2542},
    volume = {320},
    publisher = {American Medical Association},
    url = {http://jama.jamanetwork.com/article.aspx?doi=10.1001/jama.2018.19232},
    doi = {10.1001/jama.2018.19232},
    issn = {0098-7484}
}

@article{Athey2017BeyondProblems,
    title = {{Beyond prediction: Using big data for policy problems}},
    year = {2017},
    journal = {Science},
    author = {Athey, Susan},
    number = {6324},
    month = {2},
    pages = {483--485},
    volume = {355},
    publisher = {American Association for the Advancement of Science},
    url = {http://www.ncbi.nlm.nih.gov/pubmed/28154050},
    isbn = {1095-9203 (Electronic) 0036-8075 (Linking)},
    doi = {10.1126/science.aal4321},
    issn = {10959203},
    pmid = {28154050}
}

@article{Bates2014BigPatients,
    title = {{Big data in health care: Using analytics to identify and manage high-risk and high-cost patients}},
    year = {2014},
    journal = {Health Affairs},
    author = {Bates, David W. and Saria, Suchi and Ohno-Machado, Lucila and Shah, Anand and Escobar, Gabriel},
    number = {7},
    month = {8},
    pages = {1123--1131},
    volume = {33},
    publisher = {Project HOPE},
    doi = {10.1377/hlthaff.2014.0041},
    issn = {15445208},
    pmid = {25006137}
}

@article{Billings2006CasePatients,
    title = {{Case finding for patients at risk of readmission to hospital: Development of algorithm to identify high risk patients}},
    year = {2006},
    journal = {British Medical Journal},
    author = {Billings, John and Dixon, Jennifer and Mijanovich, Tod and Wennberg, David},
    number = {7563},
    month = {8},
    pages = {327--330},
    volume = {333},
    doi = {10.1136/bmj.38870.657917.AE},
    issn = {09598146}
}

@misc{Harrell2019ClassificationThinking,
    title = {{Classification vs. Prediction | Statistical Thinking}},
    year = {2019},
    author = {Harrell, Frank},
    url = {https://www.fharrell.com/post/classification/}
}

@article{Duan2019ClinicalTherapy,
    title = {{Clinical Value of Predicting Individual Treatment Effects for Intensive Blood Pressure Therapy}},
    year = {2019},
    journal = {Circulation: Cardiovascular Quality and Outcomes},
    author = {Duan, Tony and Rajpurkar, Pranav and Laird, Dillon and Ng, Andrew Y. and Basu, Sanjay},
    number = {3},
    month = {3},
    volume = {12},
    url = {https://www.ahajournals.org/doi/10.1161/CIRCOUTCOMES.118.005010},
    doi = {10.1161/CIRCOUTCOMES.118.005010}
}

@techreport{Hines2011Conditions2011,
    title = {{Conditions With the Largest Number of Adult Hospital Readmissions by Payer, 2011}},
    year = {2011},
    author = {Hines, Anika L and Barrett, Marguerite L and Jiang, H Joanna and Steiner, Claudia A},
    url = {http://www.rwjf.org/content/dam/web-assets/2011/10/medicare-hospital-readmissions-reduction-program.}
}

@article{Blake2015ConsumerExperiment,
    title = {{Consumer Heterogeneity and Paid Search Effectiveness: A Large-Scale Field Experiment}},
    year = {2015},
    journal = {Econometrica},
    author = {Blake, Thomas and Nosko, Chris and Tadelis, Steven},
    number = {1},
    month = {1},
    pages = {155--174},
    volume = {83},
    publisher = {The Econometric Society},
    doi = {10.3982/ecta12423},
    issn = {1468-0262}
}

@article{Horwitz2019CreatingTesting,
    title = {{Creating a Learning Health System through Rapid-Cycle, Randomized Testing}},
    year = {2019},
    journal = {New England Journal of Medicine},
    author = {Horwitz, Leora I. and Kuznetsova, Masha and Jones, Simon A.},
    number = {12},
    month = {9},
    pages = {1175--1179},
    volume = {381},
    publisher = {Massachusetts Medical Society},
    url = {http://www.nejm.org/doi/10.1056/NEJMsb1900856},
    doi = {10.1056/NEJMsb1900856},
    issn = {0028-4793}
}

@article{Bayati2014Data-drivenStudy.,
    title = {{Data-driven decisions for reducing readmissions for heart failure: general methodology and case study.}},
    year = {2014},
    journal = {PloS one},
    author = {Bayati, Mohsen and Braverman, Mark and Gillam, Michael and Mack, Karen M and Ruiz, George and Smith, Mark S and Horvitz, Eric},
    number = {10},
    pages = {e109264},
    volume = {9},
    publisher = {Public Library of Science},
    url = {http://www.ncbi.nlm.nih.gov/pubmed/25295524 http://www.pubmedcentral.nih.gov/articlerender.fcgi?artid=PMC4190088},
    doi = {10.1371/journal.pone.0109264},
    issn = {1932-6203},
    pmid = {25295524}
}

@article{Cook2014DyingUnit,
    title = {{Dying with Dignity in the Intensive Care Unit}},
    year = {2014},
    journal = {New England Journal of Medicine},
    author = {Cook, Deborah and Rocker, Graeme},
    number = {26},
    month = {6},
    pages = {2506--2514},
    volume = {370},
    publisher = {Massachussetts Medical Society},
    url = {http://www.nejm.org/doi/10.1056/NEJMra1208795},
    doi = {10.1056/NEJMra1208795},
    issn = {0028-4793}
}

@article{Howie2013EarlyImplications,
    title = {{Early palliative care in cancer treatment: Rationale, evidence and clinical implications}},
    year = {2013},
    journal = {Therapeutic Advances in Medical Oncology},
    author = {Howie, Lynn and Peppercorn, Jeffrey},
    number = {6},
    pages = {318--323},
    volume = {5},
    publisher = {SAGE Publications},
    doi = {10.1177/1758834013500375},
    issn = {17588359}
}

@article{Rubin1974EstimatingStudies,
    title = {{Estimating causal effects of treatments in randomized and nonrandomized studies}},
    year = {1974},
    journal = {Journal of Educational Psychology},
    author = {Rubin, Donald B.},
    number = {5},
    month = {10},
    pages = {688--701},
    volume = {66},
    doi = {10.1037/h0037350},
    issn = {00220663}
}

@article{Rubin2006EstimatingMethodology,
    title = {{Estimating the Causal Effects of Marketing Interventions Using Propensity Score Methodology}},
    year = {2006},
    journal = {Statistical Science},
    author = {Rubin, Donald B. and Waterman, Richard P.},
    pages = {206--222},
    volume = {21},
    publisher = {Institute of Mathematical Statistics},
    url = {https://www.jstor.org/stable/27645750},
    doi = {10.2307/27645750}
}

@techreport{AtheyEstimatingApplication,
    title = {{Estimating Treatment Effects with Causal Forests: An Application}},
    author = {Athey, Susan and Wager, Stefan},
    arxivId = {1902.07409v1}
}

@article{Wager2018EstimationForests,
    title = {{Estimation and Inference of Heterogeneous Treatment Effects using Random Forests}},
    year = {2018},
    journal = {Journal of the American Statistical Association},
    author = {Wager, Stefan and Athey, Susan},
    pages = {1--15},
    volume = {1459},
    publisher = {Taylor {\&} Francis},
    url = {https://doi.org/10.1080/01621459.2017.1319839},
    isbn = {0027-8424, 1091-6490},
    doi = {10.1080/01621459.2017.1319839},
    issn = {1537274X},
    pmid = {27382149},
    arxivId = {1510.04342}
}

@article{McWilliams2017FocusingCosts,
    title = {{Focusing on High-Cost Patients — The Key to Addressing High Costs?}},
    year = {2017},
    journal = {New England Journal of Medicine},
    author = {McWilliams, J. Michael and Schwartz, Aaron L.},
    number = {9},
    month = {3},
    pages = {807--809},
    volume = {376},
    publisher = {Massachussetts Medical Society},
    url = {http://www.nejm.org/doi/10.1056/NEJMp1612779},
    doi = {10.1056/NEJMp1612779},
    issn = {0028-4793}
}

@article{Athey2019GeneralizedForests,
    title = {{Generalized Random Forests}},
    year = {2019},
    journal = {The Annals of Statistics},
    author = {Athey, Susan and Tibshirani, Julie and Wager, Stefan},
    number = {2},
    pages = {1148--1178},
    volume = {47},
    url = {https://doi.org/10.1214/18-AOS1709},
    doi = {10.1214/18-AOS1709}
}

@article{Finkelstein2020HealthTrial,
    title = {{Health Care Hotspotting — A Randomized, Controlled Trial}},
    year = {2020},
    journal = {New England Journal of Medicine},
    author = {Finkelstein, Amy and Zhou, Annetta and Taubman, Sarah and Doyle, Joseph},
    number = {2},
    month = {1},
    pages = {152--162},
    volume = {382},
    publisher = {Massachussetts Medical Society},
    url = {http://www.nejm.org/doi/10.1056/NEJMsa1906848},
    doi = {10.1056/NEJMsa1906848},
    issn = {0028-4793}
}

@article{Kleinberg2018HumanPredictions,
    title = {{Human Decisions and Machine Predictions}},
    year = {2018},
    journal = {Quarterly Journal of Economics},
    author = {Kleinberg, Jon and Lakkaraju, Himabindu and Leskovec, Jure and Ludwig, Jens and Mullainathan, Sendhil},
    number = {1},
    month = {2},
    pages = {237--293},
    volume = {133},
    url = {http://www.ncbi.nlm.nih.gov/pubmed/29755141 http://www.pubmedcentral.nih.gov/articlerender.fcgi?artid=PMC5947971},
    doi = {10.1093/qje/qjx032},
    issn = {0033-5533},
    pmid = {29755141}
}

@article{Freund2011IdentificationPrograms,
    title = {{Identification of patients likely to benefit from care management programs}},
    year = {2011},
    journal = {American Journal of Managed Care},
    author = {Freund, Tobias and Mahler, Cornelia and Erler, Antje and Gensichen, Jochen and Ose, Dominik and Szecsenyi, Joachim and Peters-Klimm, Frank},
    number = {5},
    month = {5},
    pages = {345--352},
    volume = {17},
    issn = {10880224},
    pmid = {21718082}
}

@article{Weissman2011IdentifyingCare.,
    title = {{Identifying patients in need of a palliative care assessment in the hospital setting: a consensus report from the Center to Advance Palliative Care.}},
    year = {2011},
    journal = {Journal of palliative medicine},
    author = {Weissman, David E. and Meier, Diane E.},
    number = {1},
    pages = {17--23},
    volume = {14},
    doi = {10.1089/jpm.2010.0347},
    issn = {15577740}
}

@article{Goldfield2008IdentifyingReadmissions,
    title = {{Identifying potentially preventable readmissions}},
    year = {2008},
    journal = {Health Care Financing Review},
    author = {Goldfield, Norbert I. and McCullough, Elizabeth C. and Hughes, John S. and Tang, Ana M. and Eastman, Beth and Rawlins, Lisa K. and Averill, Richard F.},
    number = {1},
    month = {9},
    pages = {75--91},
    volume = {30},
    publisher = {Centers for Medicare and Medicaid Services},
    issn = {01958631},
    pmid = {19040175}
}

@article{Avati2018ImprovingLearning,
    title = {{Improving palliative care with deep learning}},
    year = {2018},
    journal = {BMC Medical Informatics and Decision Making},
    author = {Avati, Anand and Jung, Kenneth and Harman, Stephanie and Downing, Lance and Ng, Andrew and Shah, Nigam H.},
    number = {S4},
    month = {12},
    pages = {122},
    volume = {18},
    publisher = {BioMed Central},
    url = {https://bmcmedinformdecismak.biomedcentral.com/articles/10.1186/s12911-018-0677-8},
    doi = {10.1186/s12911-018-0677-8},
    issn = {1472-6947}
}

@article{Fihn2014InsightsAdministration,
    title = {{Insights From Advanced Analytics At The Veterans Health Administration}},
    year = {2014},
    journal = {Health Affairs},
    author = {Fihn, Stephan D. and Francis, Joseph and Clancy, Carolyn and Nielson, Christopher and Nelson, Karin and Rumsfeld, John and Cullen, Theresa and Bates, Jack and Graham, Gail L.},
    number = {7},
    month = {7},
    pages = {1203--1211},
    volume = {33},
    publisher = { Health Affairs },
    url = {http://www.healthaffairs.org/doi/10.1377/hlthaff.2014.0054},
    doi = {10.1377/hlthaff.2014.0054},
    issn = {0278-2715}
}

@article{Hansen2011InterventionsReview,
    title = {{Interventions to reduce 30-day rehospitalization: A systematic review}},
    year = {2011},
    journal = {Annals of Internal Medicine},
    author = {Hansen, Luke O. and Young, Robert S. and Hinami, Keiki and Leung, Alicia and Williams, Mark V.},
    number = {8},
    pages = {520--528},
    volume = {155},
    doi = {10.7326/0003-4819-155-8-201110180-00008},
    pmid = {22007045}
}

@article{Chen2017MachineExpectations,
    title = {{Machine Learning and Prediction in Medicine — Beyond the Peak of Inflated Expectations}},
    year = {2017},
    journal = {New England Journal of Medicine},
    author = {Chen, Jonathan H. and Asch, Steven M.},
    number = {26},
    month = {6},
    pages = {2507--2509},
    volume = {376},
    publisher = {Massachusetts Medical Society},
    url = {http://www.nejm.org/doi/10.1056/NEJMp1702071},
    doi = {10.1056/NEJMp1702071},
    issn = {0028-4793}
}

@article{MedPAC2016MedicareSystem,
    title = {{Medicare Advantage Program Payment System}},
    year = {2016},
    author = {{MedPAC}},
    url = {http://www.medpac.gov/docs/default-source/payment-basics/medpac_payment_basics_16_ma_final.pdf}
}

@article{Kunzel2019MetalearnersLearning,
    title = {{Metalearners for estimating heterogeneous treatment effects using machine learning}},
    year = {2019},
    journal = {Proceedings of the National Academy of Sciences},
    author = {K{\"{u}}nzel, Sören R. and Sekhon, Jasjeet S. and Bickel, Peter J. and Yu, Bin},
    number = {10},
    month = {3},
    pages = {4156--4165},
    volume = {116},
    publisher = {National Academy of Sciences},
    url = {https://www.pnas.org/content/116/10/4156},
    doi = {10.1073/PNAS.1804597116},
    issn = {0027-8424},
    pmid = {30770453}
}

@article{Escobar2019MultiyearSystem,
    title = {{Multiyear Rehospitalization Rates and Hospital Outcomes in an Integrated Health Care System}},
    year = {2019},
    journal = {JAMA Network Open},
    author = {Escobar, Gabriel J. and Plimier, Colleen and Greene, John D. and Liu, Vincent and Kipnis, Patricia},
    number = {12},
    month = {12},
    pages = {e1916769},
    volume = {2},
    url = {https://jamanetwork.com/journals/jamanetworkopen/fullarticle/2756259},
    doi = {10.1001/jamanetworkopen.2019.16769},
    issn = {2574-3805}
}

@article{Escobar2015NonelectiveMortality,
    title = {{Nonelective Rehospitalizations and Postdischarge Mortality}},
    year = {2015},
    journal = {Medical Care},
    author = {Escobar, Gabriel J. and Ragins, Arona and Scheirer, Peter and Liu, Vincent and Robles, Jay and Kipnis, Patricia},
    number = {11},
    month = {11},
    pages = {916--923},
    volume = {53},
    url = {http://www.ncbi.nlm.nih.gov/pubmed/26465120 http://www.pubmedcentral.nih.gov/articlerender.fcgi?artid=PMC4605276 http://content.wkhealth.com/linkback/openurl?sid=WKPTLP:landingpage&an=00005650-201511000-00002},
    doi = {10.1097/MLR.0000000000000435},
    issn = {0025-7079},
    pmid = {26465120}
}

@article{Walkey2020NovelInitiative,
    title = {{Novel tools for a learning health system: A combined difference-in-difference/regression discontinuity approach to evaluate effectiveness of a readmission reduction initiative}},
    year = {2020},
    journal = {BMJ Quality and Safety},
    author = {Walkey, Allan J. and Bor, Jacob and Cordella, Nicholas J.},
    number = {2},
    month = {2},
    pages = {161--167},
    volume = {29},
    publisher = {BMJ Publishing Group},
    url = {http://www.ncbi.nlm.nih.gov/pubmed/31843880},
    doi = {10.1136/bmjqs-2019-009734},
    issn = {20445415}
}

@inproceedings{Kohavi2013OnlineScale,
    title = {{Online controlled experiments at large scale}},
    year = {2013},
    booktitle = {Proceedings of the ACM SIGKDD International Conference on Knowledge Discovery and Data Mining},
    author = {Kohavi, Ron and Deng, Alex and Frasca, Brian and Walker, Toby and Xu, Ya and Pohlmann, Nils},
    month = {8},
    pages = {1168--1176},
    volume = {Part F128815},
    publisher = {Association for Computing Machinery},
    url = {http://dl.acm.org/citation.cfm?doid=2487575.2488217},
    address = {New York, New York, USA},
    isbn = {9781450321747},
    doi = {10.1145/2487575.2488217}
}

@article{Finkelstein2015OversightResearch,
    title = {{Oversight on the borderline: Quality improvement and pragmatic research}},
    year = {2015},
    journal = {Clinical Trials},
    author = {Finkelstein, Jonathan A. and Brickman, Andrew L. and Capron, Alexander and Ford, Daniel E. and Gombosev, Adrijana and Greene, Sarah M. and Iafrate, R. Peter and Kolaczkowski, Laura and Pallin, Sarah C. and Pletcher, Mark J. and Staman, Karen L. and Vazquez, Miguel A. and Sugarman, Jeremy},
    number = {5},
    month = {10},
    pages = {457--466},
    volume = {12},
    publisher = {SAGE Publications Ltd},
    doi = {10.1177/1740774515597682}
}

@article{Kent2018PersonalizedEffects,
    title = {{Personalized evidence based medicine: predictive approaches to heterogeneous treatment effects}},
    year = {2018},
    journal = {BMJ},
    author = {Kent, David M and Steyerberg, Ewout and van Klaveren, David},
    month = {12},
    pages = {k4245},
    volume = {363},
    publisher = {British Medical Journal Publishing Group},
    url = {http://www.ncbi.nlm.nih.gov/pubmed/30530757 http://www.bmj.com/lookup/doi/10.1136/bmj.k4245},
    doi = {10.1136/bmj.k4245},
    issn = {0959-8138},
    pmid = {30530757}
}

@article{Kleinberg2015PredictionProblems,
    title = {{Prediction Policy Problems}},
    year = {2015},
    journal = {American Economic Review: Papers {\&} Proceedings},
    author = {Kleinberg, Jon and Ludwig, Jens and Mullainathan, Sendhil and Obermeyer, Ziad},
    number = {5},
    pages = {491--495},
    volume = {105},
    url = {http://dx.doi.org/10.1257/aer.p20151023},
    doi = {10.1257/aer.p20151023}
}

@article{Einav2018PredictiveLife,
    title = {{Predictive modeling of U.S. health care spending in late life}},
    year = {2018},
    journal = {Science},
    author = {Einav, Liran and Finkelstein, Amy and Mullainathan, Sendhil and Obermeyer, Ziad},
    number = {6396},
    month = {6},
    pages = {1462--1465},
    volume = {360},
    publisher = {American Association for the Advancement of Science},
    url = {www.newyorker.com/magazine/2010/08/02/letting-go-2.},
    doi = {10.1126/science.aar5045},
    issn = {10959203}
}

@article{Auerbach2016PreventabilityPatients,
    title = {{Preventability and causes of readmissions in a national cohort of general medicine patients}},
    year = {2016},
    journal = {JAMA Internal Medicine},
    author = {Auerbach, Andrew D. and Kripalani, Sunil and Vasilevskis, Eduard E. and Sehgal, Neil and Lindenauer, Peter K. and Metlay, Joshua P. and Fletcher, Grant and Ruhnke, Gregory W. and Flanders, Scott A. and Kim, Christopher and Williams, Mark V. and Thomas, Larissa and Giang, Vernon and Herzig, Shoshana J. and Patel, Kanan and Boscardin, W. John and Robinson, Edmondo J. and Schnipper, Jeffrey L.},
    number = {4},
    month = {4},
    pages = {484--493},
    volume = {176},
    publisher = {American Medical Association},
    doi = {10.1001/jamainternmed.2015.7863},
    issn = {21686106}
}

@article{Leppin2014PreventingTrials,
    title = {{Preventing 30-day hospital readmissions: A systematic review and meta-analysis of randomized trials}},
    year = {2014},
    journal = {JAMA Internal Medicine},
    author = {Leppin, Aaron L. and Gionfriddo, Michael R. and Kessler, Maya and Brito, Juan Pablo and Mair, Frances S. and Gallacher, Katie and Wang, Zhen and Erwin, Patricia J. and Sylvester, Tanya and Boehmer, Kasey and Ting, Henry H. and Murad, M. Hassan and Shippee, Nathan D. and Montori, Victor M.},
    number = {7},
    month = {7},
    pages = {1095--1107},
    volume = {174},
    publisher = {American Medical Association},
    url = {http://archinte.jamanetwork.com/article.aspx?doi=10.1001/jamainternmed.2014.1608},
    doi = {10.1001/jamainternmed.2014.1608}
}

@article{Flaks-Manov2020PreventingPrediction,
    title = {{Preventing Hospital Readmissions: Healthcare Providers’ Perspectives on “Impactibility” Beyond EHR 30-Day Readmission Risk Prediction}},
    year = {2020},
    journal = {Journal of General Internal Medicine},
    author = {Flaks-Manov, Natalie and Srulovici, Einav and Yahalom, Rina and Perry-Mezre, Henia and Balicer, Ran and Shadmi, Efrat},
    month = {3},
    pages = {1--6},
    publisher = {Springer},
    doi = {10.1007/s11606-020-05739-9},
    pmid = {32141041}
}

@article{Steventon2017PreventingRisk,
    title = {{Preventing hospital readmissions: The importance of considering 'impactibility,' not just predicted risk}},
    year = {2017},
    journal = {BMJ Quality and Safety},
    author = {Steventon, Adam and Billings, John},
    number = {10},
    month = {10},
    pages = {782--785},
    volume = {26},
    publisher = {BMJ Publishing Group},
    doi = {10.1136/bmjqs-2017-006629}
}

@article{Rich1993PreventionStudy.,
    title = {{Prevention of readmission in elderly patients with congestive heart failure: results of a prospective, randomized pilot study.}},
    year = {1993},
    journal = {Journal of General Internal Medicine},
    author = {Rich, M W and Vinson, J M and Sperry, J C and Shah, A S and Spinner, L R and Chung, M K and Davila-Roman, V},
    number = {11},
    month = {11},
    pages = {585--90},
    volume = {8},
    url = {http://www.ncbi.nlm.nih.gov/pubmed/8289096},
    doi = {10.1007/bf02599709},
    pmid = {8289096}
}

@article{Breiman2001RandomForests,
    title = {{Random Forests}},
    year = {2001},
    journal = {Machine Learning},
    author = {Breiman, Leo},
    number = {1},
    month = {11},
    pages = {5--32},
    volume = {45},
    url = {http://link.springer.com/article/10.1023/A%3A1010933404324 http://link.springer.com/content/pdf/10.1023%2FA%3A1010933404324.pdf http://link.springer.com/article/10.1023%2FA%3A1010933404324},
    doi = {10.1023/A:1010933404324},
    issn = {0885-6125, 1573-0565},
    language = {en}
}

@article{Roland2012ReducingTrack,
    title = {{Reducing emergency admissions: are we on the right track?}},
    year = {2012},
    journal = {BMJ},
    author = {Roland, M. and Abel, G.},
    number = {sep18 1},
    month = {9},
    pages = {e6017-e6017},
    volume = {345},
    url = {http://www.bmj.com/cgi/doi/10.1136/bmj.e6017},
    doi = {10.1136/bmj.e6017},
    issn = {1756-1833}
}

@article{Jencks2009RehospitalizationsProgram,
    title = {{Rehospitalizations among Patients in the Medicare Fee-for-Service Program}},
    year = {2009},
    journal = {New England Journal of Medicine},
    author = {Jencks, Stephen F. and Williams, Mark V. and Coleman, Eric A.},
    number = {14},
    month = {4},
    pages = {1418--1428},
    volume = {360},
    url = {http://www.ncbi.nlm.nih.gov/pubmed/19339721 http://www.nejm.org/doi/abs/10.1056/NEJMsa0803563},
    doi = {10.1056/NEJMsa0803563},
    issn = {0028-4793},
    pmid = {19339721}
}

@article{Ascarza2018RetentionIneffective,
    title = {{Retention Futility: Targeting High-Risk Customers Might be Ineffective}},
    year = {2018},
    journal = {Journal of Marketing Research},
    author = {Ascarza, Eva},
    number = {1},
    month = {2},
    pages = {80--98},
    volume = {55},
    publisher = {American Marketing Association},
    url = {http://journals.sagepub.com/doi/10.1509/jmr.16.0163},
    doi = {10.1509/jmr.16.0163},
    issn = {0022-2437}
}

@article{Kent2016RiskTrials.,
    title = {{Risk and treatment effect heterogeneity: re-analysis of individual participant data from 32 large clinical trials.}},
    year = {2016},
    journal = {International journal of epidemiology},
    author = {Kent, David M and Nelson, Jason and Dahabreh, Issa J and Rothwell, Peter M and Altman, Douglas G and Hayward, Rodney A},
    number = {6},
    pages = {2075--2088},
    volume = {45},
    publisher = {Oxford University Press},
    url = {http://www.ncbi.nlm.nih.gov/pubmed/27375287 http://www.pubmedcentral.nih.gov/articlerender.fcgi?artid=PMC5841614},
    doi = {10.1093/ije/dyw118},
    issn = {1464-3685},
    pmid = {27375287}
}

@article{Kansagara2011RiskReview.,
    title = {{Risk prediction models for hospital readmission: a systematic review.}},
    year = {2011},
    journal = {JAMA},
    author = {Kansagara, Devan and Englander, Honora and Salanitro, Amanda and Kagen, David and Theobald, Cecelia and Freeman, Michele and Kripalani, Sunil},
    number = {15},
    month = {10},
    pages = {1688--98},
    volume = {306},
    publisher = {NIH Public Access},
    url = {http://www.ncbi.nlm.nih.gov/pubmed/22009101 http://www.pubmedcentral.nih.gov/articlerender.fcgi?artid=PMC3603349},
    doi = {10.1001/jama.2011.1515},
    issn = {1538-3598},
    pmid = {22009101}
}

@article{Rose1985SickPopulations,
    title = {{Sick Individuals and Sick Populations}},
    year = {1985},
    journal = {Internationa! Journal of Epidemiology {\textcopyright} International Epidemiological Association},
    author = {Rose, Geoffrey},
    number = {1},
    volume = {14},
    url = {https://academic.oup.com/ije/article-abstract/14/1/32/694724}
}

@misc{Kansagara2016SoLiterature,
    title = {{So many options, where do we start? An overview of the care transitions literature}},
    year = {2016},
    booktitle = {Journal of Hospital Medicine},
    author = {Kansagara, Devan and Chiovaro, Joseph C. and Kagen, David and Jencks, Stephen and Rhyne, Kerry and O'Neil, Maya and Kondo, Karli and Relevo, Rose and Motu'apuaka, Makalapua and Freeman, Michele and Englander, Honora},
    number = {3},
    month = {3},
    pages = {221--230},
    volume = {11},
    publisher = {John Wiley and Sons Inc.},
    doi = {10.1002/jhm.2502},
    issn = {15535606}
}

@article{Baily2006SpecialSafety,
    title = {{Special Report: The Ethics of Using QI Methods to Improve Health Care Quality and Safety}},
    year = {2006},
    journal = {Hastings Center Report},
    author = {Baily, Mary Ann and Bottrell, Melissa M. and Lynn, Joanne and Jennings, Bruce},
    number = {4},
    pages = {S1-S40},
    volume = {36},
    publisher = {Johns Hopkins University Press},
    doi = {10.1353/hcr.2006.0054},
    issn = {1552-146X}
}

@article{Fonarow2017TheReconsider,
    title = {{The Hospital Readmission Reduction Program Is Associated With Fewer Readmissions, More Deaths: Time to Reconsider}},
    year = {2017},
    journal = {Journal of the American College of Cardiology},
    author = {Fonarow, Gregg C. and Konstam, Marvin A. and Yancy, Clyde W.},
    number = {15},
    month = {10},
    pages = {1931--1934},
    volume = {70},
    publisher = {Elsevier USA},
    doi = {10.1016/j.jacc.2017.08.046},
    issn = {15583597},
    pmid = {28982507}
}

@article{Gupta2018ThePolicy,
    title = {{The Hospital Readmissions Reduction Program-learning from failure of a healthcare policy}},
    year = {2018},
    journal = {European Journal of Heart Failure},
    author = {Gupta, Ankur and Fonarow, Gregg C.},
    number = {8},
    month = {8},
    pages = {1169--1174},
    volume = {20},
    publisher = {John Wiley and Sons Ltd},
    url = {http://doi.wiley.com/10.1002/ejhf.1212},
    doi = {10.1002/ejhf.1212},
    issn = {13889842}
}

@article{Liu2019TheHealthcare,
    title = {{The number needed to benefit: estimating the value of predictive analytics in healthcare}},
    year = {2019},
    journal = {Journal of the American Medical Informatics Association},
    author = {Liu, Vincent X and Bates, David W and Wiens, Jenna and Shah, Nigam H},
    number = {6},
    volume = {13},
    publisher = {Oxford University Press (OUP)},
    doi = {10.1093/jamia/ocz088}
}

@article{Hansen2008TheScore,
    title = {{The prognostic analogue of the propensity score}},
    year = {2008},
    journal = {Biometrika},
    author = {Hansen, Ben B.},
    number = {2},
    month = {2},
    pages = {481--488},
    volume = {95},
    publisher = {Narnia},
    url = {https://academic.oup.com/biomet/article-lookup/doi/10.1093/biomet/asn004},
    doi = {10.1093/biomet/asn004},
    issn = {00063444}
}

@article{vanKlaveren2018TheEffects,
    title = {{The proposed ‘concordance-statistic for benefit’ provided a useful metric when modeling heterogeneous treatment effects}},
    year = {2018},
    journal = {Journal of Clinical Epidemiology},
    author = {van Klaveren, David and Steyerberg, Ewout W. and Serruys, Patrick W. and Kent, David M.},
    month = {2},
    pages = {59--68},
    volume = {94},
    publisher = {Elsevier USA},
    doi = {10.1016/j.jclinepi.2017.10.021},
    issn = {18785921}
}

@article{Chiolero2015TheStrategy,
    title = {{The pseudo-high-risk prevention strategy}},
    year = {2015},
    journal = {International Journal of Epidemiology},
    author = {Chiolero, Arnaud and Paradis, Gilles and Paccaud, Fred},
    number = {5},
    pages = {1469--1473},
    volume = {44},
    url = {https://academic.oup.com/ije/article-abstract/44/5/1469/2594556},
    doi = {10.1093/ije/dyv102}
}

@article{Sprenger2017ThreeMeasures,
    title = {{Three Arguments for Absolute Outcome Measures}},
    year = {2017},
    journal = {Philosophy of Science},
    author = {Sprenger, Jan and Stegenga, Jacob},
    pages = {840--852},
    volume = {84},
    url = {http://www.journals.uchicago.edu/t-and-c},
    isbn = {202021:16:29}
}

@article{HealthITAnalytics2018TopHttps://healthitanalytics.com/news/top-4-big-data-analytics-strategies-to-reduce-hospital-readmissions,
    title = {{Top 4 Big Data Analytics Strategies to Reduce Hospital Readmissions. https://healthitanalytics.com/news/top-4-big-data-analytics-strategies-to-reduce-hospital-readmissions}},
    year = {2018},
    author = {{Health IT Analytics}},
    url = {https://healthitanalytics.com/news/top-4-big-data-analytics-strategies-to-reduce-hospital-readmissions}
}

@article{Lindquist2011UnderstandingFactors,
    title = {{Understanding preventable hospital readmissions: Masqueraders, markers, and true causal factors}},
    year = {2011},
    journal = {Journal of Hospital Medicine},
    author = {Lindquist, Lee A. and Baker, David W.},
    number = {2},
    month = {2},
    pages = {51--53},
    volume = {6},
    doi = {10.1002/jhm.901},
    issn = {15535592}
}

@article{HealthITAnalytics2016UsingHealthitanalytics.com/features/using-risk-scores-stratification-for-population-health-management,
    title = {{Using Risk Scores, Stratification for Population Health Management. healthitanalytics.com/features/using-risk-scores-stratification-for-population-health-management}},
    year = {2016},
    author = {{Health IT Analytics}},
    url = {https://healthitanalytics.com/features/using-risk-scores-stratification-for-population-health-management}
}

\end{spacing}

\end{document}